\documentclass[11pt]{article}
\newtheorem{lemma}{Lemma}
\newtheorem{proposition}{Proposition}

\newtheorem{remark}{Remark}
\usepackage[toc,page]{appendix}
\newenvironment{proof}{\noindent{\it Proof.}}{\hfill\fbox{}\vspace*{1mm}}
\usepackage[usenames]{color}

\usepackage{amssymb}
\usepackage{amsfonts}
\usepackage{amsmath}
\usepackage{graphicx}
\usepackage{bm}
\usepackage{float}
\usepackage{booktabs}
\usepackage{tikz}
\usepackage{pgfplots}
\usetikzlibrary{positioning}
\usepackage{caption}
\usepackage{subcaption}
\usepackage{tabularx}
\usepackage{adjustbox}
\usepackage{multirow}

\definecolor{Gray}{gray}{0.9}
\definecolor{BarBlue}{RGB}{91, 124, 153}
\definecolor{BarGreen}{RGB}{143, 188, 143}
\definecolor{BarOrange}{RGB}{240, 128, 128}
\definecolor{BarPurple}{RGB}{229,196,148}
\definecolor{BarYellow}{RGB}{190,174,212}

\RequirePackage[normalem]{ulem} 
 
\providecommand{\DIFdeltex}[1]{{\protect\color{red}\sout{#1}}}                     
 
\usepackage{xcolor}
\usepackage{changebar}
\newif\ifdiff
\difftrue
\ifdiff
  \newcommand{\del}[1]{\DIFdeltex{#1}}

\else
  \newcommand{\del}[1]{}

\fi

\begin{document}
\title{\bf Fill Probabilities in a Limit Order Book with State-Dependent Stochastic Order Flows}

\author{Felix Lokin
\thanks{ Delft Institute of Applied Mathematics, TU Delft, 2628 CD Delft, The Netherlands.
(E-mail: fjp.lokin@gmail.com).}
\and Fenghui Yu
\thanks{ Delft Institute of Applied Mathematics, TU Delft, 2628 CD Delft, The Netherlands.
(E-mail: fenghui.yu@tudelft.nl).}}

\maketitle

\begin{abstract}

This paper studies the fill probabilities of limit orders placed at different price levels in a limit order book. These probabilities play a central role in execution optimization, as limit orders are not guaranteed to be executed and inherently involve a trade-off between execution cost and execution risk. We model the limit order book within a general state-dependent stochastic framework, representing its dynamics as a collection of interacting queuing systems while incorporating key stylized market features. Within this framework, we derive semi-analytical expressions for several quantities of interest under state-dependent order flows, including the probability of a mid-price change, the fill probabilities of orders placed at the best quotes, and those of orders placed deeper in the book before the opposite best quote moves. While the framework can be extended to even deeper price levels, the corresponding fill probabilities are typically negligible. We validate the proposed model through extensive numerical experiments using real foreign exchange spot market data. The results demonstrate that the model remains tractable while capturing essential order book dynamics, and that the derived expressions achieve good accuracy in estimating fill probabilities.
\end{abstract}


\section{Introduction}

Recent technological advances have made algorithmic trading a dominant force in modern financial markets, particularly in highly liquid environments. Algorithmic systems are now routinely used for tasks ranging from signal generation and market making to execution optimization and transaction cost reduction. In this paper, we focus on a key ingredient of execution optimization: the fill probability of a limit order. Since limit orders are not guaranteed to be executed, accurate estimates of fill probabilities are essential for managing execution risk and for designing trading strategies that balance execution certainty against price improvement.

The fill probability (or execution probability) of a limit order is the probability that the order is executed within a given time horizon. It depends on order-specific characteristics, such as the price level and queue position, as well as on prevailing market conditions at the time of submission. Estimating fill probabilities is challenging due to the high-frequency and rapidly evolving nature of limit order books. Nevertheless, reliable estimation is crucial in practice. In particular, traders must account for fill probabilities when choosing between aggressive strategies, which typically achieve faster execution at less favorable prices, and passive strategies, which offer better prices but carry a higher risk of non-execution.

A central trade-off in limit order placement is therefore between execution speed and price improvement. Orders submitted at or near the best quotes are more likely to be filled, since market orders interact first with the best available prices. By contrast, orders placed deeper in the book offer more favorable prices but face substantially lower execution likelihood. This trade-off motivates the need for tractable and accurate models of fill probabilities, which can be incorporated into execution and market-making algorithms.

Several approaches for estimating fill probabilities have been proposed in the literature. A commonly used simplification in optimal trading problems assumes that fill probabilities decay exponentially with the distance from the best quote. This assumption yields mathematically tractable execution models and has been widely adopted in the optimal execution literature, including Cartea et al.~(2015), Cartea and Jaimungal~(2015), and Gu\'eant et al.~(2012). While convenient, such reduced-form specifications do not explicitly account for time dependence or the evolving state of the order book.

Another strand of research relies on econometric estimation from historical data. Regression-based approaches relate execution likelihood to explanatory variables such as order size, spread, volatility, and queue imbalance. Survival analysis has been particularly influential, as it provides a natural framework for modeling time-to-fill distributions. For example, Cho and Nelling~(2000) model market order arrivals using a non-homogeneous Poisson process and assume a Weibull distribution for time-to-fill. Lo et al.~(2002) similarly apply survival analysis to compute execution-time distributions conditional on market covariates. These models are interpretable and relatively easy to estimate, but they may struggle to capture the full complexity of high-frequency order book dynamics.

More recently, machine learning methods have been proposed to estimate fill probabilities in high-dimensional and non-linear settings. These approaches can incorporate rich feature sets and adapt to changing market conditions. For instance, Maglaras et al.~(2022), Fabre and Ragel~(2023), and Arroyo et al.~(2024) combine machine learning with survival analysis to estimate both fill probabilities and fill times. Despite their flexibility, these methods typically require large training datasets, entail substantial computational costs, and often suffer from limited interpretability.

Stochastic order book models form another important class of approaches. These models explicitly describe the random evolution of order arrivals, cancellations, and executions, and allow fill probabilities to be derived analytically or computed via simulation. Smith et al.~(2003) show that even simplified stochastic models can reproduce key stylized facts of limit order books. In particular, Cont et al.~(2010) and Huang et al.~(2015) model the limit order book as a queueing system, where each price level is represented by a birth--death process: limit order arrivals correspond to ``births'', while market orders and cancellations correspond to ``deaths''. Cont et al.~(2010) further assume that order flow intensities are deterministic functions of the distance to the opposite best quote. These models are analytically tractable and can be calibrated directly from order book data, but they often rely on restrictive assumptions in order to remain manageable.

In this paper, we adopt a stochastic modeling approach and develop a general state-dependent order flow framework for estimating fill probabilities. In contrast to models with deterministic intensities, we allow the arrival and cancellation rates of limit and market orders to depend on a vector of stylized state variables, which may include queue sizes, the spread, and other observable order book features. This specification is motivated by empirical evidence that order flow intensities vary significantly with market conditions and order book states. While our framework is formulated in general terms, we also present several explicit model specifications, including classical parametric models from the literature and the model employed in our numerical experiments.

Within this framework, we model the limit order book as a system of interacting state-dependent queues, and characterize fill times in terms of first-passage times of associated birth--death and pure-death processes. By exploiting Laplace-transform techniques, we derive semi-analytical expressions for several conditional probabilities of direct relevance to execution. These include the probability of an upward or downward mid-price movement, the fill probabilities of orders placed at the best bid and best ask before the mid-price changes, and the fill probabilities of orders placed one price level deeper than the best quote before the opposite best quote moves.

A key contribution of this paper is that we go beyond the best bid and ask levels studied in much of the existing analytical literature (e.g., Cont et al.~2010). We provide a unified approach for deriving fill probabilities at deeper levels of the order book under general state-dependent intensities. While the methodology extends to arbitrarily deep levels, the resulting expressions become increasingly complex. Moreover, our empirical analysis of (Foreign Exchange)FX spot market data suggests that fill probabilities beyond one tick from the best quote are typically negligible. For this reason, we focus on the most practically relevant cases: orders posted at the best quote and one level deeper.

The main contributions of this paper can be summarized as follows:
\begin{itemize}
   \item We develop a flexible state-dependent stochastic order flow framework for limit order books, modeling limit order arrivals, market order arrivals, and cancellations through an interacting queueing system.

    \item Within this framework, we derive tractable semi-analytical expressions for key execution quantities, including the probabilities of upward and downward mid-price movements and the fill probabilities of limit orders posted at the best and second-best quotes, under general state-dependent order flow intensities.

    \item We provide, to the best of our knowledge, the first analytical characterization of fill probabilities for orders posted one price level deeper than the best quote.
    \item We validate the framework through extensive numerical experiments calibrated to real FX spot limit order book data, demonstrating both tractability and good empirical accuracy.
\end{itemize}

The remainder of the paper is organized as follows. In Section~\ref{chaptermodel}, we introduce the state-dependent stochastic order flow model and present three explicit intensity specifications. Section~\ref{chapter: mathematicaltools} reviews the mathematical tools required for our analysis, including state-dependent first-passage times of birth--death and pure-death processes. Section~\ref{Ch: FillProbTheory} derives tractable expressions for mid-price movement probabilities and fill probabilities at the best quotes and one level deeper. In Section~\ref{sec: inv_LT_comparison}, we describe the calibration procedure and present an extensive empirical and numerical study using FX spot market data. Section~\ref{conclusion} concludes the paper.

\section{Limit Order Book Model}\label{chaptermodel}
\subsection{Limit Order Book as a Stochastic Model}\label{LOBmodel}

Following the framework of Cont et al.~(2010) and Huang et al.~(2015), we model a limit order book in which participants submit limit orders at discrete price levels, each corresponding to an integer multiple of the tick size. These price levels are represented by the grid $\{1,\dots,N\}$, where $N$ is chosen sufficiently large so that the probability of orders being placed beyond level $N$ is negligible over the time horizon considered.

The state of the order book is described by the continuous-time process
\begin{equation}
\boldsymbol{Q}(t) = (Q_1(t), \dots, Q_N(t))_{t\geq 0},
\end{equation}
where $|Q_i(t)|$ denotes the number of outstanding limit orders at price level $i$ at time $t$, for $i=1,\dots,N$. To distinguish between the bid and ask sides, we adopt the convention that bid queues are represented by negative values. In particular, $Q_i(t)<0$ corresponds to $-Q_i(t)$ outstanding bid orders at level $i$, whereas $Q_i(t)>0$ corresponds to $Q_i(t)$ outstanding ask orders.

The best ask price $p_A(t)$ is defined as the lowest level containing outstanding sell orders, namely
\begin{equation}
p_A(t) := \inf\{i\in\{1,\dots,N\}: Q_i(t)>0\}\wedge (N+1).
\end{equation}
Similarly, the best bid price $p_B(t)$ is the highest level containing outstanding buy orders,
\begin{equation}
p_B(t) := \sup\{i\in\{1,\dots,N\}: Q_i(t)<0\}\vee 0.
\end{equation}

The mid-price $p_M(t)$ and the bid--ask spread $S(t)$ are then given by
\begin{equation}
p_M(t) := \frac{p_B(t)+p_A(t)}{2},
\qquad
S(t) := p_A(t)-p_B(t).
\end{equation}

\subsection{Dynamics of State-Dependent Order Flows}\label{orderflow}
The limit order book dynamics are modeled as a queuing system driven by three types of events:
\begin{itemize}
    \item limit order arrivals,
    \item market order arrivals,
    \item limit order cancellations.
\end{itemize}

Let $\boldsymbol{Q}(t_j)$ denote the state of the order book immediately before the $j$-th event, and $\boldsymbol{Q}(t_{j+1})$ the state immediately after the next event at time $t_{j+1}$. We assume that all orders have unit size (corresponding to the average order size in the data) and that events occur sequentially. Queue sizes on the bid side are represented by negative values, while queue sizes on the ask side are positive. The order book evolves according to the following update rules:
\begin{equation}\label{eq:update_rules}
\renewcommand{\arraystretch}{1.3}
\begin{array}{lll}
\toprule
\text{Event} & \text{Condition} & \text{Queue update} \\ 
\midrule
\text{Limit buy arrival}  & i<p_A   & Q_i(t_{j+1}) = Q_i(t_j)-1 \\ 
\text{Limit sell arrival} & i>p_B   & Q_i(t_{j+1}) = Q_i(t_j)+1 \\[0.3em]
\text{Buy market order}   & i=p_A   & Q_{p_A}(t_{j+1}) = Q_{p_A}(t_j)-1 \\ 
\text{Sell market order}  & i=p_B   & Q_{p_B}(t_{j+1}) = Q_{p_B}(t_j)+1 \\[0.3em]
\text{Bid cancellation}   & i\le p_B & Q_i(t_{j+1}) = Q_i(t_j)+1 \\ 
\text{Ask cancellation}   & i\ge p_A & Q_i(t_{j+1}) = Q_i(t_j)-1 \\ 
\bottomrule
\end{array}
\end{equation}

Following standard practice in modeling order flows, we assume that arrivals and cancellations are governed by independent Poisson processes with exponentially distributed inter-arrival times. Specifically, for each price level $i\in\{1,\ldots,N\}$, we assume that
\begin{itemize}
    \item limit orders arrive at rate $\lambda_{Q_i}(\boldsymbol{X}_i)$,
    \item market orders arrive at rate $\mu_{Q_i}(\boldsymbol{X}_i)$ for $i\in\{p_A,p_B\}$,
    \item limit order cancellations occur at rate $\phi_{Q_i}(\boldsymbol{X}_i)$,
\end{itemize}
where $\boldsymbol{X}_i$ is a vector of stylized state variables affecting the intensities. All rates are allowed to be state-dependent.

This framework allows for a flexible specification of the local state $\boldsymbol{X}_i$ and accommodates relevant market-specific features. We distinguish the queue size $Q_i$ from the factor vector $\boldsymbol{X}_i$ to emphasize the direct role of queue sizes in driving the order book evolution and the associated fill probabilities, while also simplifying subsequent derivations.

For notational convenience, we write
\[
\lambda_{Q_i}:=\lambda_{Q_i}(\boldsymbol{X}_i), 
\qquad 
\mu_{Q_i}:=\mu_{Q_i}(\boldsymbol{X}_i), 
\qquad 
\phi_{Q_i}:=\phi_{Q_i}(\boldsymbol{X}_i).
\]
Schematic illustrations of the dynamics at the best bid and best ask are provided in Figures~\ref{SchematicBid} and~\ref{SchematicAsk}.

\begin{figure}[H]
    \centering
    \begin{tikzpicture}[
        roundnodeb/.style={circle, draw=black!60, fill=black!10, thick, minimum size=14mm},
        roundnodew/.style={circle, minimum size=14mm}]

        
        \node[roundnodew]        (1round)                                 {$\cdots$};
        \node[roundnodeb]        (2round)        [right =1.5cm of 1round] {$Q-1$};
        \node[roundnodeb]        (3round)        [right =1.5cm of 2round] {$Q$};
        \node[roundnodeb]        (4round)        [right =1.5cm of 3round] {$Q+1$};
        \node[roundnodew]        (5round)        [right =1.5cm of 4round] {$\cdots$};
        \node[roundnodeb]        (6round)        [right =1.5cm of 5round] {$0$};

        \path [-stealth]
        (1round.north) edge [bend left] node[above] {$\mu_{\ldots}+\phi_{\ldots}$}  (2round.north)
        (2round.north) edge [bend left] node[above] {$\mu_{Q-1}+\phi_{Q-1}$}  (3round.north)
        (3round.north) edge [bend left] node[above] {$\mu_{Q}+\phi_{Q}$}  (4round.north)
        (4round.north) edge [bend left] node[above] {$\mu_{Q+1}+\phi_{Q+1}$}  (5round.north)
        (5round.north) edge [bend left] node[above] {$\mu_{\ldots}+\phi_{\ldots}$}  (6round.north)
        
        (2round.south) edge [bend left] node[below] {$\lambda_{Q-1}$}  (1round.south)
        (3round.south) edge [bend left] node[below] {$\lambda_Q$}  (2round.south)
        (4round.south) edge [bend left] node[below] {$\lambda_{Q+1}$}  (3round.south)
        (5round.south) edge [bend left] node[below] {$\lambda_{\ldots}$}  (4round.south)
        (6round.south) edge [bend left] node[below] {$\lambda_0$}  (5round.south);

    \end{tikzpicture}
    \caption[Schematic representation of the order book dynamics at the best bid.]{Schematic representation of the order book dynamics at the best bid $Q= Q_{p_B}$.}
    \label{SchematicBid}
\end{figure}
\begin{figure}[H]
    \centering
    \begin{tikzpicture}[
        roundnodeb/.style={circle, draw=black!60, fill=black!10, thick, minimum size=14mm},
        roundnodew/.style={circle, minimum size=14mm}]

        
        \node[roundnodeb]        (1round)                                 {$0$};
        \node[roundnodew]        (2round)        [right =1.5cm of 1round] {$\cdots$};
        \node[roundnodeb]        (3round)        [right =1.5cm of 2round] {$Q-1$};
        \node[roundnodeb]        (4round)        [right =1.5cm of 3round] {$Q$};
        \node[roundnodeb]        (5round)        [right =1.5cm of 4round] {$Q+1$};
        \node[roundnodew]        (6round)        [right =1.5cm of 5round] {$\cdots$};

        \path [-stealth]
        (2round.south) edge [bend left] node[below] {$\mu_{\ldots}+\phi_{\ldots}$}  (1round.south)
        (3round.south) edge [bend left] node[below] {$\mu_{Q-1}+\phi_{Q-1}$}  (2round.south)
        (4round.south) edge [bend left] node[below] {$\mu_{Q}+\phi_{Q}$}  (3round.south)
        (5round.south) edge [bend left] node[below] {$\mu_{Q+1}+\phi_{Q+1}$}  (4round.south)
        (6round.south) edge [bend left] node[below] {$\mu_{\ldots}+\phi_{\ldots}$}  (5round.south)
        
        (1round.north) edge [bend left] node[above] {$\lambda_0$}  (2round.north)
        (2round.north) edge [bend left] node[above] {$\lambda_{\ldots}$}  (3round.north)
        (3round.north) edge [bend left] node[above] {$\lambda_{Q-1}$}  (4round.north)
        (4round.north) edge [bend left] node[above] {$\lambda_Q$}  (5round.north)
        (5round.north) edge [bend left] node[above] {$\lambda_{Q+1}$}  (6round.north);   
        
    \end{tikzpicture}
    \caption[Schematic representation of the order book dynamics at the best ask.]{Schematic representation of the order book dynamics at the best ask $Q= Q_{p_A}$.}
    \label{SchematicAsk}
\end{figure}

\subsection{Examples of Order Flow Models}\label{examplesofmodels}
The intensity functions governing the arrival and cancellation of limit and market orders can be specified flexibly to reflect the characteristics of the market under consideration. In particular, the stylized factor vector $\boldsymbol{X}_i$ associated with each price level $i$ may incorporate relevant state variables such as queue sizes, the spread, or the distance to the best quotes. Below we present three representative specifications that fit naturally within our general framework.

\subsubsection{Model I}

Cont et al.~(2010) propose a model in which order flow intensities are deterministic functions of the distance to the best bid and ask prices. Their specification can be viewed as a special case of our framework. At time $t_j$, the intensities at price level $i$ are given by
\begin{equation}
\left\{
\begin{array}{ll}
\lambda_{Q_i}(\boldsymbol{X}_i) = \dfrac{\beta}{(p_A(t_j) - i)^\alpha}, 
& \text{for } i < p_A(t_j), \\[10pt]
\lambda_{Q_i}(\boldsymbol{X}_i) = \dfrac{\beta}{(i - p_B(t_j))^\alpha}, 
& \text{for } i > p_B(t_j), \\[10pt]
\mu_{Q_i}(\boldsymbol{X}_i) = \mu, 
& \text{for } i = p_A(t_j) \text{ or } i = p_B(t_j), \\[10pt]
\phi_{Q_i}(\boldsymbol{X}_i) = \theta(i - p_B(t_j))\,|Q_i(t_j)|, 
& \text{for } i \geq p_A(t_j), \\[10pt]
\phi_{Q_i}(\boldsymbol{X}_i) = \theta(p_A(t_j) - i)\,|Q_i(t_j)|, 
& \text{for } i \leq p_B(t_j),
\end{array}
\right.
\end{equation}
where $\alpha,\beta,\mu>0$ are constants and $\theta:\mathbb{N}\to\mathbb{R}_+$ is a non-negative function of the distance to the opposite best quote. In this case, one may take
\[
\boldsymbol{X}_i(t_j)=(p_A(t_j)-i,\; i-p_B(t_j),\; Q_i(t_j)).
\]
The parameters $\alpha$ and $\beta$ are typically calibrated via least-squares fitting.

\subsubsection{Model II}

Motivated by the empirical study of Toke and Yoshida (2017), one may alternatively model order flow intensities using a parametric specification that depends on the spread and queue size. At time $t_j$, the intensities at level $i$ are given by
\begin{equation} \left\{ \begin{array}{ll} \lambda_{Q_i}(\boldsymbol{X}_i) = & \exp\bigg[\alpha_0 + \alpha_1 \log(S(t_j)) + \alpha_{11} \log^2(S(t_j)) + \alpha_2 \log(1+Q_i(t_j)) \\ &\quad + \alpha_{22} \log^2(1+Q_i(t_j)) + \alpha_{12} \log(S(t_j)) \log(1+Q_i(t_j))\bigg], \\[10pt] \mu_{Q_i}(\boldsymbol{X}_i) = & \exp\bigg[\beta_0 + \beta_1 \log(S(t_j)) + \beta_{11} \log^2(S(t_j)) + \beta_2 \log(1+Q_i(t_j)) \\ &\quad + \beta_{22} \log^2(1+Q_i(t_j)) + \beta_{12} \log(S(t_j)) \log(1+Q_i(t_j))\bigg], \\[10pt] \phi_{Q_i}(\boldsymbol{X}_i) = & \exp\bigg[\gamma_0 + \gamma_1 \log(S(t_j)) + \gamma_{11} \log^2(S(t_j)) + \gamma_2 \log(1+Q_i(t_j)) \\ &\quad + \gamma_{22} \log^2(1+Q_i(t_j)) + \gamma_{12} \log(S(t_j)) \log(1+Q_i(t_j))\bigg], \end{array} \right. \end{equation}
where $\alpha_i,\alpha_{ij},\beta_i,\beta_{ij},\gamma_i,\gamma_{ij}$ are constants for $i,j\in\mathbb{N}$. In this specification, the stylized factor vector can be taken as
\[
\boldsymbol{X}_i(t_j)=(S(t_j),\; Q_i(t_j)).
\]
Calibration may be performed via likelihood maximization; see Toke and Yoshida (2017) for further details.

\subsubsection{Model III}

For our numerical experiments based on real order book data from the FX spot market, we propose a specification inspired by Cont et al.~(2010) and supported by our empirical findings. While model selection is not the primary focus of this work, Model III provides a concrete setting in which we illustrate our methodology. Other specifications may be treated analogously.

At time $t_j$, the intensities are specified as
\begin{equation}
\left\{
\begin{array}{ll}
\lambda_{Q_i}(\boldsymbol{X}_i) = \lambda(p_A(t_j) - i, S(t_j)), 
& \text{for } i < p_A(t_j), \\[10pt]
\lambda_{Q_i}(\boldsymbol{X}_i) = \lambda(i - p_B(t_j), S(t_j)), 
& \text{for } i > p_B(t_j), \\[10pt]
\mu_{Q_i}(\boldsymbol{X}_i) = \mu(S(t_j)), 
& \text{for } i = p_A(t_j) \text{ or } i = p_B(t_j), \\[10pt]
\phi_{Q_i}(\boldsymbol{X}_i) = \theta(i - p_B(t_j), S(t_j))\,|Q_i(t_j)|, 
& \text{for } i \geq p_A(t_j), \\[10pt]
\phi_{Q_i}(\boldsymbol{X}_i) = \theta(p_A(t_j) - i, S(t_j))\,|Q_i(t_j)|, 
& \text{for } i \leq p_B(t_j),
\end{array}
\right.
\end{equation}
where $\lambda,\theta:\mathbb{N}\times\mathbb{N}\to\mathbb{R}_+$ are non-negative functions of the distance to the opposite best quote and the spread, and $\mu:\mathbb{N}\to\mathbb{R}_+$ is a non-negative function of the spread. Here, the absolute value accounts for the convention that bid-side queue sizes are negative. In this case, we take
\[
\boldsymbol{X}_i(t_j)=(p_A(t_j)-i,\; i-p_B(t_j),\; S(t_j)).
\]

The resulting order book dynamics define a continuous-time Markov process on $\mathbb{Z}^N$, with transitions occurring at order arrivals and cancellations. In particular, the transitions are as follows:
\begin{table}[h!]
\centering
\renewcommand{\arraystretch}{1.3}
\begin{tabular}{llll}
\toprule
Event & Condition & Queue update & Rate \\
\midrule
Limit buy arrival& $i<p_A(t_j)$ 
& $Q_i(t_{j+1}) = Q_i(t_j)-1$ 
& $\lambda(p_A(t_j)-i,\; S(t_j))$ \\

Limit sell arrival &$i>p_B(t_j)$ 
& $Q_i(t_{j+1}) = Q_i(t_j)+1$ 
& $\lambda(i-p_B(t_j),\; S(t_j))$ \\[0.3em]

Buy market order &$i=p_A(t_j)$ 
& $Q_{p_A}(t_{j+1}) = Q_{p_A}(t_j)-1$ 
& $\mu(S(t_j))$ \\

Sell market order &$i=p_B(t_j)$ 
& $Q_{p_B}(t_{j+1}) = Q_{p_B}(t_j)+1$ 
& $\mu(S(t_j))$ \\[0.3em]

Bid cancellation &$i\le p_B(t_j)$ 
& $Q_i(t_{j+1}) = Q_i(t_j)+1$ 
& $\theta(p_A(t_j)-i,\; S(t_j))\,|Q_i(t_j)|$ \\

Ask cancellation &$i\ge p_A(t_j)$ 
& $Q_i(t_{j+1}) = Q_i(t_j)-1$ 
& $\theta(i-p_B(t_j),\; S(t_j))\,|Q_i(t_j)|$ \\
\bottomrule
\end{tabular}
\label{tab:modelIII_transitions}
\end{table}

\section{Preliminaries on State-Dependent Queuing Systems}\label{chapter: mathematicaltools}

To monitor the continuous-time dynamics of the limit order book $\boldsymbol{Q}(t)$, a natural modeling framework is provided by queueing theory as described in Section \ref{orderflow}, where the number of outstanding orders at each price level is represented as a queue. By viewing the agent’s limit order as the last order in the corresponding queue, execution of this order occurs precisely when that queue becomes empty. Consequently, the queue containing the agent’s order can be modeled as a pure death process, since orders arriving after the agent do not directly affect its execution. In contrast, the remaining queues evolve as birth--death processes, reflecting the arrival, cancellation, and execution of orders over time.

\subsection{First-Passage Times of State-Dependent Birth--Death Processes}\label{SectionFPT}

A birth--death process is a continuous-time Markov process on $\mathbb{N}_0$ whose transitions occur only between neighboring states. From state $i$, the process jumps to $i+1$ (a ``birth'') at rate $\lambda_i$, and to $i-1$ (a ``death'') at rate $\mu_i$, where the state $i$ represents the number of units in the system. Throughout, we allow the rates to depend on the current state, and assume that $\mu_0=0$.

Let $\sigma_b$ denote the first-passage time to state $0$ starting from $b\geq 1$. Then $\sigma_b$ can be decomposed as
\begin{equation}\label{fpt_sum}
\sigma_b
=
\sigma_{b,b-1}+\sigma_{b-1,b-2}+\cdots+\sigma_{1,0},
\end{equation}
where $\sigma_{i,i-1}$ denotes the first-passage time from state $i$ to state $i-1$. By the strong Markov property, the random variables $\{\sigma_{i,i-1}\}_{i=1}^b$ are independent. Hence, if $\hat{f}_b(s)$ denotes the Laplace transform of $\sigma_b$ and $\hat{f}_{i,i-1}(s)$ denotes the Laplace transform of $\sigma_{i,i-1}$, we obtain
\begin{equation}\label{LTproduct}
\hat{f}_b(s)
=
\prod_{i=1}^b \hat{f}_{i,i-1}(s).
\end{equation}

A semi-analytical expression for $\hat{f}_{i,i-1}(s)$ can be obtained in terms of continued fractions. In particular, Abate and Whitt (1999) show that
\begin{equation}\label{LTbirth}
\hat{f}_{i,i-1}(s)
=
-\frac{1}{\lambda_{i-1}}
\text{{\huge K}}_{k=i}^\infty
\frac{-\lambda_{k-1}\mu_k}{\lambda_k+\mu_k+s},
\end{equation}
where the continued fraction is defined by
\begin{equation}\label{confrac}
\text{{\huge K}}_{n=1}^\infty \frac{a_n}{b_n}
:=
\cfrac{a_1}{b_1+\cfrac{a_2}{b_2+\cfrac{a_3}{b_3+\cdots}}},
\end{equation}
with sequences $\{a_n\},\{b_n\}\subset\mathbb{C}$ satisfying $a_n\neq 0$ for all $n$. For convenience, we also use the compact notation
\[
\text{{\huge K}}_{n=1}^\infty \frac{a_n}{b_n}
=
\frac{a_1}{b_1+}\frac{a_2}{b_2+}\frac{a_3}{b_3+}\cdots.
\]
Numerical approximation schemes for continued fractions are provided in Appendix~\ref{numerical2appendix}.

Combining~\eqref{LTproduct} and~\eqref{LTbirth}, we obtain the Laplace transform of the first-passage time $\sigma_b$ as
\begin{equation}\label{laplacef_b}
\hat{f}_b(s)
=
\prod_{i=1}^b
\left(
-\frac{1}{\lambda_{i-1}}
\text{{\huge K}}_{k=i}^\infty
\frac{-\lambda_{k-1}\mu_k}{\lambda_k+\mu_k+s}
\right).
\end{equation}
The probability density function of $\sigma_b$ can then be recovered numerically via Laplace inversion. Two inversion techniques are described in Appendix~\ref{numericalappendix}.

\subsection{First-Passage Times of State-Dependent Pure-Death Processes}\label{SectionFPTDeath}

We next consider a state-dependent pure-death process, in which only downward transitions are possible. Specifically, from state $i\geq 1$ the process jumps to $i-1$ at rate $\mu_i$. Let $\epsilon_b$ denote the first-passage time to state $0$ starting from $b\geq 1$. Then $\epsilon_b$ admits the decomposition
\begin{equation}\label{fpt_sum_death}
\epsilon_b
=
\epsilon_{b,b-1}+\epsilon_{b-1,b-2}+\cdots+\epsilon_{1,0},
\end{equation}
where $\epsilon_{i,i-1}$ denotes the first-passage time from state $i$ to $i-1$. By the strong Markov property, the random variables $\{\epsilon_{i,i-1}\}_{i=1}^b$ are independent. Hence, if $g_b(t)$ denotes the density of $\epsilon_b$ and $\hat{g}_b(s)$ its Laplace transform, we obtain
\begin{equation}\label{LTproduct_death}
\hat{g}_b(s)
=
\prod_{i=1}^b \hat{g}_{i,i-1}(s),
\end{equation}
where $\hat{g}_{i,i-1}(s)$ is the Laplace transform of the density of $\epsilon_{i,i-1}$.

Since $\epsilon_{i,i-1}$ is exponentially distributed with parameter $\mu_i$, its density is given by
\[
g_{i,i-1}(t)=\mu_i e^{-\mu_i t},
\qquad t\geq 0,
\]
and therefore
\begin{equation}\label{LT_puredeath}
\hat{g}_{i,i-1}(s)
=
\int_0^\infty \mu_i e^{-(\mu_i+s)t}\,dt
=
\frac{\mu_i}{\mu_i+s}.
\end{equation}
Substituting~\eqref{LT_puredeath} into~\eqref{LTproduct_death} yields the Laplace transform of $\epsilon_b$:
\begin{equation}\label{productdeath}
\hat{g}_b(s)
=
\prod_{i=1}^b \frac{\mu_i}{\mu_i+s}.
\end{equation}

\section{Fill Probabilities in a Limit Order Book}\label{Ch: FillProbTheory}
In this section, we study the computation of fill probabilities in a limit order book with state-dependent order flows. Here, ``state-dependent'' refers not only to the queue sizes at each price level, but also to additional stylized factors that affect the intensities of order arrivals and cancellations.

We derive expressions for the probability of a mid-price change, both upward and downward, and for the probability that a limit order placed at the best bid or best ask is executed before the mid-price moves. In addition, we obtain a semi-analytical expression for the fill probability of a limit order posted one price level deeper than the best quote, conditional on the event that the opposite best quote does not move first. While the methodology can be extended to deeper levels, the analytical expressions quickly become more involved. Moreover, empirical evidence from FX spot market data suggests that executions beyond one tick from the best quotes occur only rarely.

Our derivations rely on first-passage time analysis for state-dependent birth--death processes.

\subsection{Probability of a Change in Mid-Price}\label{sec:pricemove_theory}

In this subsection, we study the probability of the first change in the mid-price, distinguishing between an upward and a downward movement. Let $\tau$ denote the time of the first change in the mid-price, defined by
\begin{equation*}
    \tau = \inf\{t \geq 0 : p_M(t) \neq p_M(0)\},
\end{equation*}
where $p_M(t)$ denotes the mid-price at time $t$. Since the mid-price depends solely on the best bid and best ask prices, the probability that the next price move is an increase, conditional on the state of the order book at time $t=0$, is given by
\begin{equation}\label{pricemoveprob}
    \mathbb{P}\left[p_M(\tau) > p_M(0) \,\middle|\, Q_{p_A}(0) = q_0^A,\, Q_{p_B}(0) = q_0^B,\, S(0) = s_0\right],
\end{equation}
where $s_0 \geq 1$ denotes the initial spread.  
Similarly, the probability that the next price move is a decrease is
\begin{equation}\label{pricemoveprob2}
    \mathbb{P}\left[p_M(\tau) < p_M(0) \,\middle|\, Q_{p_A}(0) = q_0^A,\, Q_{p_B}(0) = q_0^B,\, S(0) = s_0\right].
\end{equation}

The following proposition provides an explicit characterization of these probabilities.

\begin{proposition}\label{proppricemove}
Let $\sigma_A$ and $\sigma_B$ denote the first-passage times to $0$ of the best ask and best bid queues, respectively. For $i\in\{A,B\}$, let $\hat{f}^{s_0}_{\sigma_i}(s)$ denote the Laplace transform of the density of $\sigma_i$, conditional on the initial spread $s_0\geq 1$. Then
\begin{equation}\label{FPTtransform}
\hat{f}^{s_0}_{\sigma_i}(s)
=
\prod_{j=1}^{q_0^i}
\left(
-\frac{1}{\lambda_{j-1}(\boldsymbol{X}_{p_i})}
\,\text{\textup{\huge K}}_{k=j}^\infty
\frac{-\lambda_{k-1}(\boldsymbol{X}_{p_i})
\left(\mu_k(\boldsymbol{X}_{p_i})+\phi_k(\boldsymbol{X}_{p_i})\right)}
{\lambda_k(\boldsymbol{X}_{p_i})+\mu_k(\boldsymbol{X}_{p_i})+\phi_k(\boldsymbol{X}_{p_i})+s}
\right).
\end{equation}

Define
\[
\Lambda_{s_0}
:=
\sum_{m=1}^{s_0-1}\lambda_0(\boldsymbol{X}_{p_A-m})
=
\sum_{m=1}^{s_0-1}\lambda_0(\boldsymbol{X}_{p_B+m}),
\]
and let $i,j\in\{A,B\}$ with $i\neq j$. Then the conditional probabilities in~\eqref{pricemoveprob} and~\eqref{pricemoveprob2} can be obtained by evaluating at $s=0$ the inverse Laplace transform of
\begin{equation}\label{cdflap}
\hat{F}^{s_0}_{\sigma_i,\sigma_j}(s)
=
\frac{1}{s}
\left(
\hat{f}^{s_0}_{\sigma_i}(\Lambda_{s_0}+s)
+
\frac{\Lambda_{s_0}}{\Lambda_{s_0}+s}
\Bigl(1-\hat{f}^{s_0}_{\sigma_i}(\Lambda_{s_0}+s)\Bigr)
\right)
\left(
\hat{f}^{s_0}_{\sigma_j}(\Lambda_{s_0}-s)
+
\frac{\Lambda_{s_0}}{\Lambda_{s_0}-s}
\Bigl(1-\hat{f}^{s_0}_{\sigma_j}(\Lambda_{s_0}-s)\Bigr)
\right).
\end{equation}
In particular, $(i,j)=(A,B)$ corresponds to an upward mid-price move, while $(i,j)=(B,A)$ corresponds to a downward mid-price move.

In the special case $s_0=1$, when no orders can be posted inside the spread,~\eqref{cdflap} simplifies to
\begin{equation}\label{LTcdfpricemove}
\hat{F}^{1}_{\sigma_i,\sigma_j}(s)
=
\frac{1}{s}\hat{f}^{1}_{\sigma_i}(s)\hat{f}^{1}_{\sigma_j}(-s).
\end{equation}
\end{proposition}

\begin{proof}
Let $\sigma_b$ denote the first-passage time to state zero for a birth-death process with state-dependent birth rates $\lambda_k$ and death rates $\mu_k$, starting from state $b$.

For each $i \in {A, B}$, let $Q_i := Q_{P_i}$. The process $Q_i$ evolves as a birth-death process with the following state-dependent transition rates:
\begin{equation}\label{birthdeathrates}
\begin{cases}
\text{birth rate} & \lambda_{Q_i}(\boldsymbol{X}_{p_i}), \\
\text{death rate} & \mu_{Q_i}(\boldsymbol{X}_{p_i}) + \phi_{Q_i}(\boldsymbol{X}_{p_i}).
\end{cases}
\end{equation}
Applying \eqref{laplacef_b} with $b = q_0^i$ yields \eqref{FPTtransform}.

Next, recall that the mid-price is defined by
$
p_M(t) := \frac{p_B(t) + p_A(t)}{2}.
$
A change in $p_M(t)$ occurs whenever $p_B$ or $p_A$ changes. A price increase happens if either $p_B(\tau) > p_B(0)$ (due to a buy limit order inside the spread) or $p_A(\tau) > p_A(0)$ (due to depletion of the best ask queue). Two cases are considered:
\begin{itemize}
\item[\textbf{(i)}] \textbf{Case \( s_0 = 1 \):}
A price move occurs only when either the best bid or best ask queue depletes. Then,
\[
\mathbb{P}[\text{price increases}] = \mathbb{P}[\sigma_A < \sigma_B] = \mathbb{P}[\sigma_A - \sigma_B < 0],
\]
and similarly,
\[
\mathbb{P}[\text{price decreases}] = \mathbb{P}[\sigma_B - \sigma_A < 0].
\]
Assuming independence, the Laplace transform of the density of $\sigma_i - \sigma_j$ is
\begin{equation}\label{pricepdf}
\hat{f}^{1}_{\sigma_i - \sigma_j}(s) = \mathbb{E}[e^{-s(\sigma_i - \sigma_j)}] = \hat{f}^{1}_{\sigma_i}(s) \hat{f}^{1}_{\sigma_j}(-s).
\end{equation}
Since the Laplace transform of a CDF $F$ is $\hat{F}(s) = \frac{1}{s} \hat{f}(s)$, we get
\begin{equation}
 \hat{F}_{\sigma_i,\sigma_j}^1(s)=\frac{1}{s}\hat{f}^1_{\sigma_i-\sigma_j}(s)=\frac{1}{s}\hat{f}_{\sigma_i}^1(s)\hat{f}_{\sigma_j}^1(-s),
\end{equation}
which proves \eqref{LTcdfpricemove}. Inversion of this transform at $s=0$ gives the desired probability.

\item[\textbf{(ii)}] \textbf{Case \( s_0 > 1 \):}
A price move can also result from a limit order posted inside the spread. Let $\tau_A^m$ (resp. $\tau_B^m$) be the first arrival time of an ask (resp. bid) order within the spread and is $m$ ticks away from the best ask (bid), for $m=1,...,s_0-1$. Each $\tau_A^m$ and $\tau_B^m$ is exponential with rate $\lambda_0(\boldsymbol{X}_{p_A - m})$ or $\lambda_0(\boldsymbol{X}_{p_B + m})$, respectively. Let $\tau_A, \tau_B$ be exponential with rate
\[
\Lambda_{s_0} = \sum_{m=1}^{s_0 - 1} \lambda_0(\boldsymbol{X}_{p_A - m}) = \sum_{m=1}^{s_0 - 1} \lambda_0(\boldsymbol{X}_{p_B + m}),
\]
representing the first arrival of a limit order inside the spread. Note that the total rate of arrival of orders posted inside the spread is the same on both sides of the book.

Then the first mid-price change occurs at
\begin{equation*}\label{probspread}
    \tau := \sigma_A \wedge \sigma_B \wedge \min \left\{ \tau_A^m, \tau_B^m : m = 1, \dots, s_0 - 1 \right\}.
\end{equation*}
A price increase occurs if either of the following events occurs first:
\begin{itemize}
    \item A buy limit order is posted inside the spread, establishing a new best bid at a higher price level;
    \item The ask queue $Q_A$ depletes before either a sell limit order is posted inside the spread or the bid queue $Q_B$ depletes.
\end{itemize}
The analysis for a price decrease is symmetric.

Then the probability in Equation~\eqref{pricemoveprob} (or \eqref{pricemoveprob2}) is equivalent to
\begin{equation}\label{increaseprobspread}
    \mathbb{P}[\sigma_i \wedge \tau_j < \sigma_j \wedge \tau_i] = \mathbb{P}[\sigma_i \wedge \tau_j - \sigma_j \wedge \tau_i < 0],
\end{equation}
where $(i, j) = (A, B)$ for a price increase and $(i, j) = (B, A)$ for a price decrease.

Since $\sigma_A$, $\sigma_B$, $\tau_A$, and $\tau_B$ are independent, we may use the same argument as in the case $s_0 = 1$. The Laplace transform of the distribution function of $\sigma_i \wedge \tau_j - \sigma_j \wedge \tau_i$ is given by
\begin{equation}\label{part1}
    \hat{F}_{\sigma_i,\sigma_j}^{s_0}(s):=\hat{F}_{{\sigma_{i}\wedge\tau_j}, {\sigma_j\wedge\tau_i}}^{s_0}(s)=\frac{1}{s}\hat{f}^{s_0}_{{\sigma_{i}\wedge\tau_j}, {\sigma_j\wedge\tau_i}}(s)=\frac{1}{s}\hat{f}_{{\sigma_{i}\wedge\tau_j}}^{s_0}(s)\hat{f}_{\sigma_j\wedge\tau_i}^{s_0}(-s).
\end{equation}

To compute $\hat{f}^{s_0}_{\sigma_i \wedge \tau_j}(s)$, we apply Lemma~\ref{LTminofexponential}, which gives the Laplace transform of the minimum of a birth-death process and an independent exponential clock. Specifically, for $i, j \in \{A, B\}$:
\begin{equation}\label{part2}
    \hat{f}^{s_0}_{\sigma_i \wedge \tau_j}(s)
    = \hat{f}^{s_0}_{\sigma_i}(\Lambda_{s_0} + s) + \frac{\Lambda_{s_0}}{\Lambda_{s_0} + s} \left(1 - \hat{f}^{s_0}_{\sigma_i}(\Lambda_{s_0} + s) \right),
\end{equation}
where $\hat{f}^{s_0}_{\sigma_i}$ is given by Equation~\eqref{FPTtransform}.

Substituting Equation~\eqref{part2} into \eqref{part1} yields the expression in Equation~\eqref{cdflap}. Hence, the Laplace transform of the distribution in \eqref{increaseprobspread} is given by \eqref{cdflap}, and the desired probabilities in \eqref{pricemoveprob} and \eqref{pricemoveprob2} can be computed by inverting this Laplace transform and evaluating it at $s = 0$, using $(i, j) = (A, B)$ and $(i, j) = (B, A)$, respectively.
\end{itemize}
\end{proof}

\begin{lemma} \label{LTminofexponential}
Let \( X \geq 0 \) be an exponential random variable with rate \( \Lambda \), and let \( Y \geq 0 \) be an independent random variable with density \( f_Y \). Denote by \( \hat{f}_Y(s) \) the Laplace transform of \( f_Y \). Then the Laplace transform of the density of \( X \wedge Y \) is given by
\begin{equation}
\hat{f}_{X \wedge Y}(s) = \hat{f}_Y(\Lambda + s) + \frac{\Lambda}{\Lambda + s} \left(1 - \hat{f}_Y(\Lambda + s) \right).
\end{equation}
\end{lemma}

\begin{proof}
See Appendix~\ref{proofappendix}.
\end{proof}

\begin{remark}\label{remark-birth-death}
Under Models I, II, and III described in Section~\ref{examplesofmodels}, the queue processes $Q_i$ have explicit transition rate functions $\lambda_{Q_i}(\boldsymbol{X}_{p_i})$, $\mu_{Q_i}(\boldsymbol{X}_{p_i})$, and $\phi_{Q_i}(\boldsymbol{X}_{p_i})$. By substituting these into Proposition~\ref{proppricemove}, we obtain closed-form expressions for the probabilities of price movements given in Equation~\eqref{pricemoveprob} and \eqref{pricemoveprob2}.

For example, under Model III, which is used in our numerical experiments, we have
\[
\Lambda_{s_0} = \sum_{m=1}^{s_0 - 1} \lambda_0(m, s_0),
\]
and the Laplace transform of the first-passage time $\sigma_i$ takes the form
\begin{equation}\label{FPTtransformmodel3}
    \hat{f}^{s_0}_{\sigma_i}(s) = \prod_{j=1}^{q^i_0} \left( -\frac{1}{\lambda_{j-1}(0, s_0)} \, \text{\textup{\huge K}}_{k=j}^\infty \frac{-\lambda_{k-1}(0, s_0) \left( \mu_k(s_0) +  \theta(0, s_0) k \right)}{\lambda_k(0, s_0) + \mu_k(s_0) + \theta(0, s_0) k + s} \right),
\end{equation}
which is used in the computations of Proposition~\ref{proppricemove}.

Furthermore, if the transition rates $\lambda_k$ and $\mu_k$ are independent of the queue state $k$, then we can simplify $\Lambda_{s_0} = \sum_{m=1}^{s_0 - 1} \lambda(m, s_0)$ and obtain
\begin{equation}\label{FPTtransformmodel3simple}
    \hat{f}^{s_0}_{\sigma_i}(s) = \prod_{j=1}^{q^i_0} \left( -\frac{1}{\lambda(0, s_0)} \, \text{\textup{\huge K}}_{k=j}^\infty \frac{-\lambda(0, s_0) \left( \mu(s_0) + \theta(0, s_0) k \right)}{\lambda(0, s_0) + \mu(s_0) + \theta(0, s_0) k + s} \right).
\end{equation}
\end{remark}

To support numerical computation, we provide two methods for Laplace transform inversion: the Euler method and the COS method. Implementation details can be found in Appendix~\ref{numericalappendix}.

\subsection{Fill Probability at the Best Quotes}\label{sec: fill_prob_best_p}

We now study the probability that a limit order posted at the best bid or best ask is executed before the mid-price changes, under the assumption that the order is never cancelled. Let $NC_A$ and $NC_B$ denote the events that an order is submitted at time $t=0$ at the best ask and best bid, respectively, and remains active thereafter.

The corresponding conditional fill probability is
\begin{equation}\label{probexecution}
\mathbb{P}\left[\epsilon_i < \tau \,\middle|\, Q_A(0) = q_0^A,\, Q_B(0) = q_0^B,\, S(0) = s_0,\, NC_i\right],
\end{equation}
where $i\in\{A,B\}$, $\tau$ denotes the first time the mid-price moves, and $\epsilon_i$ is the first-passage time to zero of the queue position of the order on price leavel $i$, initialized at $q_0^i$. In particular, $i=A$ corresponds to an order posted at the best ask, while $i=B$ corresponds to an order posted at the best bid.

To simplify notation, we omit the conditioning on the initial state $(Q_A(0)=q_0^A,\; Q_B(0)=q_0^B,\; S(0)=s_0)$ in what follows.

We model the order's queue position as a pure-death process rather than a birth--death process. This is justified by the time-priority rule: after the order is submitted, any subsequent arrivals at the same price level join the queue behind it and therefore do not affect its execution. Under this representation, the fill probability can be computed explicitly, as shown in the following proposition.

\begin{proposition}\label{prop_fillprob}
Fix $s_0\geq 1$ and let $\epsilon_i$ denote the first-passage time to zero of the queue position of a limit order posted at the best quote on side $i\in\{A,B\}$, with initial position $q_0^i\geq 1$. Then the Laplace transform of the density of $\epsilon_i$ is given by
\begin{equation}\label{Laplacedeath}
\hat{g}_{\epsilon_i}^{s_0}(s)
=
\prod_{k=1}^{q_0^i}
\frac{\mu_k(\boldsymbol{X}_{p_i})+\phi_k(\boldsymbol{X}_{p_i})}
{\mu_k(\boldsymbol{X}_{p_i})+\phi_k(\boldsymbol{X}_{p_i})+s},
\end{equation}
where $\mu_k(\boldsymbol{X}_{p_i})$ and $\phi_k(\boldsymbol{X}_{p_i})$ denote the market order and cancellation intensities at the best quote $p_i$.

Let $\sigma_j$ be the first-passage time to zero of the opposite best-quote queue, and let $\hat{f}_{\sigma_j}^{s_0}(s)$ denote the Laplace transform of the density of $\sigma_j$, as defined in~\eqref{FPTtransform}. Moreover, define
\[
\Lambda_{s_0}
=
\sum_{m=1}^{s_0-1}\lambda_0(\boldsymbol{X}_{p_A-m})
=
\sum_{m=1}^{s_0-1}\lambda_0(\boldsymbol{X}_{p_B+m}).
\]
Then the fill probability in~\eqref{probexecution} is obtained by evaluating at $s=0$ the inverse Laplace transform of
\begin{equation}\label{LTexecution}
\hat{F}_{\epsilon_i,\sigma_j}^{s_0}(s)
=
\frac{1}{s}\hat{g}_{\epsilon_i}^{s_0}(s)
\left(
\hat{f}_{\sigma_j}^{s_0}(2\Lambda_{s_0}-s)
+
\frac{2\Lambda_{s_0}}{2\Lambda_{s_0}-s}
\Bigl(1-\hat{f}_{\sigma_j}^{s_0}(2\Lambda_{s_0}-s)\Bigr)
\right),
\end{equation}
for $i\neq j\in\{A,B\}$.

In the special case $s_0=1$ (when no orders can be posted inside the spread), this simplifies to
\begin{equation}
\hat{F}_{\epsilon_i,\sigma_j}^{1}(s)
=
\frac{1}{s}\hat{g}_{\epsilon_i}^{1}(s)\,
\hat{f}_{\sigma_j}^{1}(-s).
\end{equation}
\end{proposition}

\begin{proof}
Let $\epsilon_b$ denote the first-passage time to zero of a pure-death process starting at state $b$ with death rates $\mu_k$ in state $k$. As shown in Section~\ref{SectionFPTDeath}, the Laplace transform of the density of $\epsilon_b$, denoted $\hat{g}_{\epsilon_b}(s)$, is given by Equation~\eqref{productdeath}. 

In our setting, the death rate of the pure-death process, denoted $D_i$, is given by $\mu_{D_i}(\boldsymbol{X}_{p_i}) + \phi_{D_i}(\boldsymbol{X}_{p_i})$, where $i \in \{A, B\}$. Applying Equation~\eqref{productdeath} yields
\begin{equation}
    \hat{g}_{\epsilon_i}^{s_0}(s) = \prod_{k=1}^{q_0^i} \frac{\mu_k(\boldsymbol{X}_{p_i}) + \phi_k(\boldsymbol{X}_{p_i})}{\mu_k(\boldsymbol{X}_{p_i}) + \phi_k(\boldsymbol{X}_{p_i}) + s}.
\end{equation}

We now distinguish two cases:

\begin{enumerate}
    \item[\textbf{(i)}] \textbf{Case \( s_0 = 1 \):} In this case, the mid-price cannot move through placement of new limit orders, and the fill probability reduces to
    \[
        \mathbb{P}[\epsilon_i < \sigma_j] = \mathbb{P}[\epsilon_i - \sigma_j < 0],
    \]
    for \( i \neq j \in \{A, B\} \). The Laplace transform of the density of \( \epsilon_i - \sigma_j \) is given by
    \begin{equation}\label{fillprob_pdf}
        \hat{f}_{\epsilon_i - \sigma_j}^1(s) = \mathbb{E}\left[e^{-s(\epsilon_i - \sigma_j)}\right] = \mathbb{E}[e^{-s \epsilon_i}] \, \mathbb{E}[e^{s \sigma_j}] = \hat{g}_{\epsilon_i}^1(s) \hat{f}_{\sigma_j}^1(-s).
    \end{equation}
    Hence, the Laplace transform of the CDF \( F_{\epsilon_i, \sigma_j}^1 \) of \( \epsilon_i - \sigma_j \) is
    \[
        \hat{F}_{\epsilon_i, \sigma_j}^1(s) = \frac{1}{s} \hat{f}_{\epsilon_i - \sigma_j}^1(s) = \frac{1}{s} \hat{g}_{\epsilon_i}^1(s) \hat{f}_{\sigma_j}^1(-s).
    \]
    Therefore, the desired probability is
    \[
        \mathbb{P}[\epsilon_i < \sigma_j] = F_{\epsilon_i, \sigma_j}^1(0),
    \]
    which can be obtained by evaluating the inverse Laplace transform of \( \hat{F}_{\epsilon_i, \sigma_j}^1(s) \) at $s=0$.

    \item[\textbf{(ii)}] \textbf{Case \( s_0 > 1 \):} The fill probability becomes
    \[
        \mathbb{P}[\epsilon_i < \sigma_j \wedge \tau_A \wedge \tau_B] = \mathbb{P}[\epsilon_i - (\sigma_j \wedge \tau_A \wedge \tau_B) < 0],
    \]
    where \( \tau_A \) and \( \tau_B \) are independent exponential random variables with rate \( \Lambda_{s_0} \), defined as in Proposition~\ref{proppricemove}. The minimum \( \tau_A \wedge \tau_B \) is exponentially distributed with rate \( 2\Lambda_{s_0} \).

    Let \( X = \tau_A \wedge \tau_B \sim \text{Exp}(2\Lambda_{s_0}) \) and \( Y = \sigma_j \geq 0 \). By Lemma~\ref{LTminofexponential}, the Laplace transform of the PDF of \( \sigma_j \wedge \tau_A \wedge \tau_B \) is
    \begin{equation}\label{equa:mini3}
        \hat{f}_{\sigma_j \wedge \tau_A \wedge \tau_B}^{s_0}(s) = \hat{f}_{\sigma_j}^{s_0}(2\Lambda_{s_0} - s) + \frac{2\Lambda_{s_0}}{2\Lambda_{s_0} - s} \left(1 - \hat{f}_{\sigma_j}^{s_0}(2\Lambda_{s_0} - s)\right).
    \end{equation}

    Let \( F_{\epsilon_i, \sigma_j}^{s_0} \) denote the CDF of \( \epsilon_i - (\sigma_j \wedge \tau_A \wedge \tau_B) \). The Laplace transform of this CDF is
    \begin{equation}
        \hat{F}_{\epsilon_i, \sigma_j}^{s_0}(s) = \frac{1}{s} \hat{g}_{\epsilon_i}^{s_0}(s) \hat{f}_{\sigma_j \wedge \tau_A \wedge \tau_B}^{s_0}(-s),
    \end{equation}
    which corresponds to Equation~\eqref{LTexecution} with $\hat{f}_{\sigma_j \wedge \tau_A \wedge \tau_B}^{s_0}(\cdot)$ given by \eqref{equa:mini3}. The desired fill probability is obtained by evaluating the inverse Laplace transform of \( \hat{F}_{\epsilon_i, \sigma_j}^{s_0}(s) \) at zero.
\end{enumerate}
\end{proof}

\begin{remark}\label{remark-pure-death}
Under Models I, II, and III, as described in Section~\ref{examplesofmodels}, explicit expressions for the state-dependent death rates \( \mu_{D_i}(\boldsymbol{X}_{p_i}) \) and \( \phi_{D_i}(\boldsymbol{X}_{p_i}) \) are available and are required for computing Equation~\eqref{Laplacedeath} in Proposition~\ref{prop_fillprob}.

For example, under Model III, we have
\begin{equation}\label{FP1transformmodel3}
    \hat{g}_{\epsilon_i}^{s_0}(s) = \prod_{k=1}^{q_0^i} \frac{\mu_k(s_0) + \theta(0, s_0)\,k}{\mu_k(s_0) + \theta(0, s_0)\,k + s}.
\end{equation}

Moreover, if the death rate \( \mu_k \) is independent of the queue state \( k \), i.e., \( \mu_k(s_0) = \mu(s_0) \) for all \( k \), the expression simplifies to
\begin{equation}\label{FP1transformmodel3simple}
    \hat{g}_{\epsilon_i}^{s_0}(s) = \prod_{k=1}^{q_0^i} \frac{\mu(s_0) + \theta(0, s_0)\,k}{\mu(s_0) + \theta(0, s_0)\,k + s}.
\end{equation}
\end{remark}

\subsection{Fill Probability at a Price Level Deeper than the Best Quotes}\label{sec: fill_prob_level2}

In this section, we compute the fill probability of a limit order submitted one price level deeper than the best quote, before the opposite best quote moves. More precisely, we consider buy limit orders posted at $p_B-1$ and sell limit orders posted at $p_A+1$, and focus on execution events occurring prior to a movement of the opposite best quote.

Although the methodology developed below can in principle be extended to limit orders posted deeper in the book, both the notation and the resulting analytical expressions quickly become considerably more involved. Moreover, our numerical experiments suggest that fill probabilities beyond one tick from the best quote are typically negligible. Consistent with this observation, Section~\ref{sec: empirical_execution} shows empirically that approximately $85\%$ of executed limit orders are posted within one tick of the best quote, indicating that deeper levels contribute only marginally in practice.

Under our modeling assumptions, such an order may be executed through two possible mechanisms:
\begin{enumerate}
    \item a large market order consumes liquidity across multiple price levels;
    \item the best quote queue is depleted, so that the best quote moves by one tick toward the order and the order becomes part of the new best quote.
\end{enumerate}
Since we assume unit-sized orders, the first mechanism cannot occur. Consequently, execution of an order posted one level deeper is only possible if the best quote shifts toward that level. Once this happens, the order becomes part of the best quote, and its fill probability can be computed using the approach developed in Section~\ref{sec: fill_prob_best_p}.

\medskip
We now introduce notation to formalize this argument. For $i\in\{A,B\}$, let $Q_i(t)$ denote the queue size at the best quote on side $i$ at time $t$, and let $Q_{i-}(t)$ denote the queue size one level deeper. Moreover, let $W_{i-}(t)$ denote the number of remaining orders at that deeper level that were already present at time $0$.

Let $\tau_i^{\text{quote}}$ be the first time the best quote on side $i$ moves by one tick toward the deeper level:
\begin{equation}
\tau_i^{\text{quote}} \equiv 
\begin{cases}
\inf\{t \geq 0 : p_A(t) > p_A(0)\}, & \text{if } i = A, \\[0.3em]
\inf\{t \geq 0 : p_B(t) < p_B(0)\}, & \text{if } i = B.
\end{cases}
\end{equation}
Similarly, define $\tau_i^{\text{other}}$ as the first time the mid-price changes due to any other event, such as the opposite best quote moving or the best quote moving in the opposite direction:
\begin{equation}
\tau_i^{\text{other}} \equiv 
\begin{cases}
\inf\left\{t \geq 0 : \bigl(p_A(t) < p_A(0)\bigr) \lor \bigl(p_B(t) \neq p_B(0)\bigr)\right\}, 
& \text{if } i = A, \\[0.6em]
\inf\left\{t \geq 0 : \bigl(p_B(t) > p_B(0)\bigr) \lor \bigl(p_A(t) \neq p_A(0)\bigr)\right\}, 
& \text{if } i = B.
\end{cases}
\end{equation}

The probability that the best quote moves toward the order before any such competing event occurs is
\begin{equation}\label{prob_level2_term1}
\mathbb{P}\left[\tau_i^{\text{quote}} < \tau_i^{\text{other}} \,\middle|\,
Q_A(0) = q_0^A,\, Q_B(0) = q_0^B,\, S(0) = s_0
\right],
\end{equation}
which is computed in Proposition~\ref{prop_1_level2}. For notational convenience, we omit the conditioning on the initial state and simply write $\mathbb{P}[\tau_i^{\text{quote}} < \tau_i^{\text{other}}]$.

\medskip
Once the best quote has shifted to the price level of the concerned order, we define the remaining time until the next mid-price change by
\[
\tau^i
\equiv
\inf\{t \geq \tau_i^{\text{quote}} : p_M(t) \neq p_M(\tau_i^{\text{quote}})\}
-\tau_i^{\text{quote}}.
\]
At time $\tau_i^{\text{quote}}$, the order becomes part of the best quote. Therefore, conditional on $\tau_i^{\text{quote}} < \tau_i^{\text{other}}$, its fill probability can be computed using Proposition~\ref{prop_fillprob}, with the initial time shifted to $\tau_i^{\text{quote}}$.

More precisely, conditional on $\tau_i^{\text{quote}} < \tau_i^{\text{other}}$, the fill probability of a limit order posted one level deeper than the best quote is given by
\begin{equation}\label{prob_level2_fill}
\mathbb{P}\!\left[
\epsilon_{i-} < \tau^i \,\middle|\,
W_{i-}(\tau_i^{\text{quote}})=q_{\tau_i^{\text{quote}}}^{i-},\,
Q_j(\tau_i^{\text{quote}})=q_{\tau_i^{\text{quote}}}^j,\,
S(\tau_i^{\text{quote}})=s_0+1,\,
NC_{i-}
\right],
\end{equation}
where $i\neq j\in\{A,B\}$ and $NC_{i-}$ denotes the event that a limit order is placed one level deeper than the best quote and remains active thereafter. The random variable $\epsilon_{i-}$ is the first-passage time to zero of a pure-death process representing the order's position in the queue at the deeper level. Since only one best quote moves by one tick at time $\tau_i^{\text{quote}}$, the spread increases from $s_0$ to $s_0+1$.

Combining the above, the fill probability of a limit order submitted one price level deeper than the best quote is therefore
\begin{equation}\label{prob_level2_fill_total}
\mathbb{P}\left[\tau_i^{\text{quote}} < \tau_i^{\text{other}}\right]\cdot
\mathbb{P}\!\left[
\epsilon_{i-} < \tau^i \,\middle|\,
W_{i-}(\tau_i^{\text{quote}})=q_{\tau_i^{\text{quote}}}^{i-},\,
Q_j(\tau_i^{\text{quote}})=q_{\tau_i^{\text{quote}}}^j,\,
S(\tau_i^{\text{quote}})=s_0+1,\,
NC_{i-}
\right].
\end{equation}

\begin{proposition}
The fill probability of a limit order submitted one price level deeper than the best quote, as defined in~\eqref{prob_level2_fill_total}, admits the representation
\begin{equation}\label{prob_level2}
\resizebox{1.0\hsize}{!}{$
\begin{aligned}
&\mathbb{P}[\tau_i^{\text{quote}}<\tau_i^{\text{other}}]\cdot\\
&\left(\sum_{m=1}^{q_0^{i-}}\sum_{n=1}^{N_j}
\mathbb{P}[\epsilon_{i-}<\tau^i \mid W_{i-}(\tau_i^{\text{quote}})=m,Q_j(\tau_i^{\text{quote}})=n]\,
\mathbb{P}[W_{i-}(\tau_i^{\text{quote}})=m]\,
\mathbb{P}[Q_j(\tau_i^{\text{quote}})=n]
\right).
\end{aligned}
$}
\end{equation}
where $i\neq j\in\{A,B\}$, and $N_{i-}$ and $N_j$ denote the admissible queue sizes at one level deeper than the best quote $p_i$ and at the opposite best quote $p_j$, respectively, at time $\tau_i^{\text{quote}}$.

Moreover, $\mathbb{P}[Q_j(\tau_i^{\text{quote}})=n]$ may be approximated using the stationary distribution of $Q_j$, while the distribution of $W_{i-}(\tau_i^{\text{quote}})$ is given by
\begin{equation}\label{prob_level2_bid}
\mathbb{P}[W_{i-}(\tau_i^{\text{quote}})=m]=
\begin{cases}
\mathbb{P}[\sigma_i<\epsilon_{q_0^{i-},\,q_0^{i-}-1}], & m=q_0^{i-}, \\[0.4em]
\mathbb{P}[\epsilon_{q_0^{i-},\,m}<\sigma_i]-\mathbb{P}[\epsilon_{q_0^{i-},\,m-1}<\sigma_i], & 1<m<q_0^{i-}, \\[0.4em]
\mathbb{P}[\epsilon_{q_0^{i-},\,1}<\sigma_i], & m=1,
\end{cases}
\end{equation}
where $\epsilon_{m,n}$ denotes the first-passage time of a pure-death process from state $m$ to state $n$, and $\sigma_i$ denotes the first-passage time at which the best-quote queue on side $i\in\{A,B\}$ is depleted.
\end{proposition}

\begin{proof}
At time $\tau_i^{\text{quote}}$, the queue sizes $W_{i-}(\tau_i^{\text{quote}})$ and $Q_j(\tau_i^{\text{quote}})$ are random. Hence, by the law of total probability, the conditional fill probability in~\eqref{prob_level2_fill} can be obtained by conditioning on all admissible values of these quantities and summing over the resulting scenarios. In particular, for $m\geq 1$ and $n\geq 1$, consider the joint event
\[
\{W_{i-}(\tau_i^{\text{quote}})=m,\; Q_j(\tau_i^{\text{quote}})=n\}.
\]
Its probability can be written as
\begin{equation}
\mathbb{P}\!\left[
W_{i-}(\tau_i^{\text{quote}})=m,\;
Q_j(\tau_i^{\text{quote}})=n
\right].
\end{equation}

By the assumed independence between the processes $W_{i-}$ and $Q_j$, we obtain
\begin{equation}\label{prob_level2_combination}
\mathbb{P}\!\left[
W_{i-}(\tau_i^{\text{quote}})=m,\;
Q_j(\tau_i^{\text{quote}})=n
\right]
=
\mathbb{P}[W_{i-}(\tau_i^{\text{quote}})=m]\,
\mathbb{P}[Q_j(\tau_i^{\text{quote}})=n].
\end{equation}
For notational simplicity, we omit conditioning on the initial state at $t=0$ and on the event
$S(\tau_i^{\text{quote}})=s_0+1$ and $NC_{i-}$, which are fixed throughout.

Conditioning on $(W_{i-}(\tau_i^{\text{quote}}),Q_j(\tau_i^{\text{quote}}))=(m,n)$ and summing over all admissible values yields
\begin{equation}\label{prob_level2_term2}
\sum_{m=1}^{N_{i-}}\sum_{n=1}^{N_j}
\mathbb{P}\!\left[
\epsilon_{i-}<\tau^i \,\middle|\,
W_{i-}(\tau_i^{\text{quote}})=m,\;
Q_j(\tau_i^{\text{quote}})=n
\right]
\mathbb{P}[W_{i-}(\tau_i^{\text{quote}})=m]\,
\mathbb{P}[Q_j(\tau_i^{\text{quote}})=n].
\end{equation}

Combining Equations~\eqref{prob_level2_term1} and~\eqref{prob_level2_term2}, we obtain the fill probability of an order placed one price level deeper than the best quote $q_i$, for $i\in\{A,B\}$, as
\[
\resizebox{1.0\hsize}{!}{$
\begin{aligned}
&\mathbb{P}[\tau_i^{\text{quote}}<\tau_i^{\text{other}}]\cdot\\
&\left(\sum_{m=1}^{q_0^{i-}}\sum_{n=1}^{N_j}
\mathbb{P}[\epsilon_{i-}<\tau^i \mid W_{i-}(\tau_i^{\text{quote}})=m,Q_j(\tau_i^{\text{quote}})=n]\,
\mathbb{P}[W_{i-}(\tau_i^{\text{quote}})=m]\,
\mathbb{P}[Q_j(\tau_i^{\text{quote}})=n]
\right),
\end{aligned}
$}
\]
which proves~\eqref{prob_level2}.

Finally, note that $W_{i-}$ evolves as a pure-death process and therefore takes values in a finite state space at time $\tau_i^{\text{quote}}$, which leads directly to the distribution given in~\eqref{prob_level2_bid}. This reflects the fact that the order of interest can only advance in the queue due to cancellations, while orders submitted later have lower priority and do not affect its execution. 

In contrast, the process $Q_j$ evolves as a birth--death process and is, in principle, unbounded. In practice, however, queue sizes at the best quote remain effectively bounded, allowing us to restrict attention to a finite range of values. Following Cont and De Larrard (2013), we estimate the distribution of $Q_j(\tau_i^{\text{quote}})$ empirically from observed queue sizes after the corresponding one-tick price move, and approximate $\mathbb{P}[Q_j(\tau_i^{\text{quote}})=n]$ using the stationary distribution of $Q_j$.
\end{proof}

\begin{remark}
Suppose that all transition rates $\lambda_{Q_l}(\boldsymbol{X}_l)$, $\mu_{Q_l}(\boldsymbol{X}_l)$, and $\phi_{Q_l}(\boldsymbol{X}_l)$ depend only on the queue size, namely
\[
\lambda_{Q_l}(\boldsymbol{X}_l)=\lambda_{Q_l}(p_l), \quad
\mu_{Q_l}(\boldsymbol{X}_l)=\mu_{Q_l}(p_l), \quad
\phi_{Q_l}(\boldsymbol{X}_l)=\phi_{Q_l}(p_l),
\]
for all $l=1,\ldots,N$, as introduced in Section~\ref{LOBmodel}. In this case, the invariant distribution of the limit order book admits an explicit form. Let $\pi(p_l)$ denote the stationary distribution of $Q_l$. By standard results (see Gross and Harris, 1998), it is given by
\begin{equation}
\pi_n(p_l)=\pi_0(p_l)\prod_{k=1}^{n}\rho_{k-1}(p_l),
\end{equation}
where
\begin{equation}
\pi_0(p_l)=\left(1+\sum_{n=1}^{\infty}\prod_{k=1}^{n}\rho_{k-1}(p_l)\right)^{-1},
\end{equation}
and
\begin{equation}
\rho_n(p_l)=\frac{\lambda_n(p_l)}{\mu_{n+1}(p_l)+\phi_{n+1}(p_l)}.
\end{equation}
Consequently, for $i\neq j\in\{A,B\}$, one may approximate
\[
\mathbb{P}[Q_j(\tau_i^{\text{quote}})=n]=\pi_n(p_j).
\]
\end{remark}

To compute the probability given in Equation (\ref{prob_level2}), we have the following propositions for the calculations of (\ref{prob_level2_term1}) and (\ref{prob_level2_term2}).

\begin{proposition}\label{prop_1_level2}
Recall that $\hat{f}^{s_0}_{\sigma_i}(s)$ denotes the Laplace transform of the density of the first-passage time $\sigma_i$ for the birth--death process $Q_i$ to hit $0$, given initial spread $s_0\geq 1$, and is given by~\eqref{FPTtransform}. Define
\[
\Lambda_{s_0}
=
\sum_{m=1}^{s_0-1}\lambda_0(\boldsymbol{X}_{p_A-m})
=
\sum_{m=1}^{s_0-1}\lambda_0(\boldsymbol{X}_{p_B+m}).
\]
Then the probability in~\eqref{prob_level2_term1} is obtained by evaluating at $s=0$ the inverse Laplace transform of
\begin{equation}
\hat{G}_{\sigma_i,\sigma_j}^{s_0}(s)
=
\frac{1}{s}\hat{f}_{\sigma_i}^{s_0}(s)
\left(
\hat{f}_{\sigma_j}^{s_0}(2\Lambda_{s_0}-s)
+
\frac{2\Lambda_{s_0}}{2\Lambda_{s_0}-s}
\bigl(1-\hat{f}_{\sigma_j}^{s_0}(2\Lambda_{s_0}-s)\bigr)
\right),
\end{equation}
for $i\neq j\in\{A,B\}$. In particular, when $s_0=1$, we obtain
\begin{equation}
\hat{G}_{\sigma_i,\sigma_j}^{1}(s)
=
\frac{1}{s}\hat{f}_{\sigma_i}^{1}(s)\hat{f}_{\sigma_j}^{1}(-s).
\end{equation}
\end{proposition}

\begin{proof}
We distinguish two cases depending on the initial spread $s_0$ again.

\begin{enumerate}
\item[\textbf{(i)}] \textbf{Case \( s_0 = 1 \):}
Since no orders can be posted inside the spread, the event $\{\tau_i^{\text{quote}}<\tau_i^{\text{other}}\}$ is equivalent to the event that the best-quote queue on side $i$ is depleted before the opposite best-quote queue. Hence, for $i\neq j\in\{A,B\}$,
\[
\mathbb{P}[\tau_i^{\text{quote}}<\tau_i^{\text{other}}]
=
\mathbb{P}[\sigma_i<\sigma_j]
=
\mathbb{P}[\sigma_i-\sigma_j<0].
\]
By independence of $\sigma_i$ and $\sigma_j$, this probability can be computed by evaluating at $t=0$ the inverse Laplace transform of
\[
\hat{G}_{\sigma_i,\sigma_j}^{1}(s)
=
\frac{1}{s}\hat{f}_{\sigma_i}^{1}(s)\hat{f}_{\sigma_j}^{1}(-s).
\]

\item[\textbf{(ii)}] \textbf{Case \( s_0 > 1 \):}
When the spread exceeds one tick, the mid-price may also move due to the arrival of limit orders inside the spread. Let $\tau_A$ and $\tau_B$ denote the first arrival times of sell and buy limit orders inside the spread, respectively. Then $\tau_A$ and $\tau_B$ are independent exponential random variables with rate $\Lambda_{s_0}$, and
\[
\mathbb{P}[\tau_i^{\text{quote}}<\tau_i^{\text{other}}]
=
\mathbb{P}[\sigma_i<\sigma_j\wedge\tau_A\wedge\tau_B].
\]
Applying Lemma~\ref{LTminofexponential} yields that this probability is obtained by evaluating at $s=0$ the inverse Laplace transform of
\[
\hat{G}_{\sigma_i,\sigma_j}^{s_0}(s)
=
\frac{1}{s}\hat{f}_{\sigma_i}^{s_0}(s)
\left(
\hat{f}_{\sigma_j}^{s_0}(2\Lambda_{s_0}-s)
+
\frac{2\Lambda_{s_0}}{2\Lambda_{s_0}-s}
\bigl(1-\hat{f}_{\sigma_j}^{s_0}(2\Lambda_{s_0}-s)\bigr)
\right).
\]
\end{enumerate}
\end{proof}

Since the best quote $p_i$ has moved by one tick to $p_{i-}$, the bid--ask spread at time $\tau_i^{\text{quote}}$ increases to $s_0+1$. We are therefore led to the following proposition, which provides an explicit expression for 
$\mathbb{P}[\epsilon_{i-}<\tau^i \mid W_{i-}(\tau_i^{\text{quote}})=m,\; Q_j(\tau_i^{\text{quote}})=n]$ 
appearing in~\eqref{prob_level2_term2}.

\begin{proposition}\label{prop_2_level2}
Fix $s_0\geq 1$ and suppose that at time $\tau_i^{\text{quote}}$ the spread equals $s_0+1$. Conditional on
\[
W_{i-}(\tau_i^{\text{quote}})=m,
\qquad
Q_j(\tau_i^{\text{quote}})=n,
\]
let $p'_A$ and $p'_B$ denote the best ask and best bid at time $\tau_i^{\text{quote}}$, and define
\[
\Lambda'_{s_0+1}
=
\sum_{r=1}^{s_0}\lambda_0(\boldsymbol{X}_{p'_A-r})
=
\sum_{r=1}^{s_0}\lambda_0(\boldsymbol{X}_{p'_B+r}).
\]
Let $\hat{f}_{\sigma_j}^{s_0+1}(s)$ denote the Laplace transform of the density of the first-passage time $\sigma_j$ of the birth--death process $Q_j$ to hit $0$, given $Q_j(\tau_i^{\text{quote}})=n$ (cf.~\eqref{FPTtransform}). Similarly, let $\hat{g}_{\epsilon_{i-}}^{s_0+1}(s)$ denote the Laplace transform of the density of the first-passage time $\epsilon_{i-}$ to hit $0$ for the pure-death process describing the queue at the new best quote $p_{i-}$, given $W_{i-}(\tau_i^{\text{quote}})=m$ (cf.~\eqref{Laplacedeath}). 

Then, for $i\neq j\in\{A,B\}$,
\[
\mathbb{P}\!\left[\epsilon_{i-}<\tau^i \,\middle|\,
W_{i-}(\tau_i^{\text{quote}})=m,\;
Q_j(\tau_i^{\text{quote}})=n
\right]
\]
is obtained by evaluating at $s=0$ the inverse Laplace transform of
\begin{equation}\label{LT_fill_level2_conditional}
\hat{F}_{\epsilon_{i-},\sigma_j}^{s_0+1}(s)
=
\frac{1}{s}\hat{g}_{\epsilon_{i-}}^{s_0+1}(s)
\left(
\hat{f}_{\sigma_j}^{s_0+1}(2\Lambda'_{s_0+1}-s)
+
\frac{2\Lambda'_{s_0+1}}{2\Lambda'_{s_0+1}-s}
\bigl(1-\hat{f}_{\sigma_j}^{s_0+1}(2\Lambda'_{s_0+1}-s)\bigr)
\right).
\end{equation}
\end{proposition}

\begin{proof}
Conditional on $W_{i-}(\tau_i^{\text{quote}})=m$, the queue position of the order at level $p_{i-}$ evolves as a pure-death process with death rates $\mu_k(\boldsymbol{X}_{p_{i-}})+\phi_k(\boldsymbol{X}_{p_{i-}})$ for $k=1,\ldots,m$. The first-passage time to $0$ is therefore the sum of $m$ independent exponential holding times, which yields~\eqref{Laplacedeath} with $q_0^i=W_{i-}(\tau_i^{\text{quote}})=m$.

Since the spread at time $\tau_i^{\text{quote}}$ equals $s_0+1\geq 2$, the mid-price may move either when the opposite best-quote queue $Q_j$ is depleted (at time $\sigma_j$) or when a limit order arrives inside the spread. The latter occurs at the minimum of two independent exponential clocks with total rate $2\Lambda'_{s_0+1}$. Applying Lemma~\ref{LTminofexponential} with $\epsilon_{i-}$ and $\sigma_j$ gives the Laplace representation~\eqref{LT_fill_level2_conditional}, which completes the proof.
\end{proof}

\begin{proposition}\label{prop_3_level2}
Let $\hat{f}_{\sigma_i}^{s_0}(s)$ denote the Laplace transform of the density of the first-passage time $\sigma_i$ of the birth--death process $Q_i$ to hit $0$, as given in~\eqref{FPTtransform}. 

Consider the pure-death process $\tilde{W}_{i-}$ on side $i\in\{A,B\}$, and let $\hat{h}_{q_0^{i-},m}^{i-,s_0}(s)$ denote the Laplace transform of the density of the first-passage time for $\tilde{W}_{i-}$ to move from state $q_0^{i-}$ to state $m\leq q_0^{i-}$. Then, for $m\geq 1$,
\begin{equation}
\hat{h}_{q_0^{i-},m}^{i-,s_0}(s)
=
\prod_{k=m}^{q_0^{i-}}
\frac{\phi_k(\boldsymbol{X}_{p_{i-}})}
{\phi_k(\boldsymbol{X}_{p_{i-}})+s}.
\end{equation}

Moreover, the probabilities in~\eqref{prob_level2_bid} can be obtained by evaluating at $s=0$ the inverse Laplace transform of
\begin{equation}
\hat{H}^{s_0}_{\sigma_i; q_0^{i-}, m}(s)=
\begin{cases}
\dfrac{1}{s}\hat{f}^{s_0}_{\sigma_i}(s)\,
\hat{h}^{i-, s_0}_{q_0^{i-},\,q_0^{i-}-1}(-s),
& m=q_0^{i-}, \\[0.6em]
\dfrac{1}{s}\hat{f}^{s_0}_{\sigma_i}(-s)\,
\Bigl(\hat{h}^{i-, s_0}_{q_0^{i-},\,m}(s)
-\hat{h}^{i-, s_0}_{q_0^{i-},\,m-1}(s)\Bigr),
& 1<m<q_0^{i-}, \\[0.6em]
\dfrac{1}{s}\hat{f}^{s_0}_{\sigma_i}(-s)\,
\hat{h}^{i-, s_0}_{q_0^{i-},\,1}(s),
& m=1.
\end{cases}
\end{equation}
\end{proposition}

\begin{proof}
Recall that $W_{i-}(t)$, for $0\leq t\leq \tau_i^{\text{quote}}$, denotes the number of orders at level $p_{i-}$ that were already present at time $0$ and remain outstanding up to time $t$. Since $p_{i-}$ is not the best quote prior to $\tau_i^{\text{quote}}$, these orders cannot be executed by market orders under the unit-size assumption. Hence, their evolution is driven solely by cancellations, and the process $\tilde{W}_{i-}$ is a pure-death process with death rates $\phi_k(\boldsymbol{X}_{p_{i-}})$.

By~\eqref{productdeath}, the Laplace transform of the density of the first-passage time from $q_0^{i-}$ to $m\leq q_0^{i-}$ satisfies
\[
\hat{h}_{q_0^{i-},m}^{i-,s_0}(s)
=
\prod_{k=m}^{q_0^{i-}}
\hat{h}_{k,k-1}^{i-,s_0}(s),
\qquad
\hat{h}_{k,k-1}^{i-,s_0}(s)
=
\frac{\phi_k(\boldsymbol{X}_{p_{i-}})}{\phi_k(\boldsymbol{X}_{p_{i-}})+s},
\]
which yields
\[
\hat{h}_{q_0^{i-},m}^{i-,s_0}(s)
=
\prod_{k=m}^{q_0^{i-}}
\frac{\phi_k(\boldsymbol{X}_{p_{i-}})}
{\phi_k(\boldsymbol{X}_{p_{i-}})+s}.
\]

Next, to compute the probabilities in~\eqref{prob_level2_bid}, note that on the time interval $[0,\tau_i^{\text{quote}}]$ no limit orders can be posted inside the spread. Therefore, the value of $W_{i-}(\tau_i^{\text{quote}})$ is determined by whether the pure-death process $\tilde{W}_{i-}$ reaches a given level before the depletion time $\sigma_i$ of the best-quote queue.

\begin{enumerate}
\item[\textbf{(i)}]\textbf{Case $m=q_0^{i-}$.}
We have
\[
\mathbb{P}[W_{i-}(\tau_i^{\text{quote}})=q_0^{i-}]
=
\mathbb{P}[\sigma_i<\epsilon_{q_0^{i-},\,q_0^{i-}-1}],
\]
and by independence and the Laplace-transform argument used in Proposition~\ref{prop_fillprob},
this probability is obtained by evaluating at $s=0$ the inverse Laplace transform of
\[
\hat{H}^{s_0}_{\sigma_i; q_0^{i-}, q_0^{i-}}(s)
=
\frac{1}{s}\hat{f}^{s_0}_{\sigma_i}(s)\,
\hat{h}^{i-, s_0}_{q_0^{i-},\,q_0^{i-}-1}(-s).
\]

\item[\textbf{(ii)}]\textbf{Case $1<m<q_0^{i-}$.}
In this case,
\[
\mathbb{P}[W_{i-}(\tau_i^{\text{quote}})=m]
=
\mathbb{P}[\epsilon_{q_0^{i-},\,m}<\sigma_i]
-
\mathbb{P}[\epsilon_{q_0^{i-},\,m-1}<\sigma_i],
\]
and thus it is obtained by evaluating at $s=0$ the inverse Laplace transform of
\[
\hat{H}^{s_0}_{\sigma_i; q_0^{i-}, m}(s)
=
\frac{1}{s}\hat{f}^{s_0}_{\sigma_i}(-s)
\Bigl(
\hat{h}^{i-, s_0}_{q_0^{i-},\,m}(s)
-
\hat{h}^{i-, s_0}_{q_0^{i-},\,m-1}(s)
\Bigr).
\]

\item[\textbf{(iii)}]\textbf{Case $m=1$.}
Finally,
\[
\mathbb{P}[W_{i-}(\tau_i^{\text{quote}})=1]
=
\mathbb{P}[\epsilon_{q_0^{i-},\,1}<\sigma_i],
\]
which is obtained by evaluating at $s=0$ the inverse Laplace transform of
\[
\hat{H}^{s_0}_{\sigma_i; q_0^{i-}, 1}(s)
=
\frac{1}{s}\hat{f}^{s_0}_{\sigma_i}(-s)\,
\hat{h}^{i-, s_0}_{q_0^{i-},\,1}(s).
\]
\end{enumerate}
\end{proof}

Using Propositions~\ref{prop_1_level2}--\ref{prop_3_level2}, we can compute all terms in~\eqref{prob_level2} and thereby obtain the fill probability of a limit order posted one price level deeper than the best quote, before the opposite best quote moves.

\begin{remark}
In line with Remarks~\ref{remark-birth-death} and~\ref{remark-pure-death}, once explicit expressions for the transition rates
$\lambda_{Q_i}(\boldsymbol{X}_{p_i})$, $\mu_{Q_i}(\boldsymbol{X}_{p_i})$, $\phi_{Q_i}(\boldsymbol{X}_{p_i})$,
and $\mu_{Q_{i-}}(\boldsymbol{X}_{p_{i-}})$, $\phi_{Q_{i-}}(\boldsymbol{X}_{p_{i-}})$ for $i\in\{A,B\}$ are specified, the corresponding fill probability in~\eqref{prob_level2} follows directly from Propositions~\ref{prop_1_level2}--\ref{prop_3_level2}.
\end{remark}

\section{Numerical and Empirical Experiments}\label{sec: inv_LT_comparison}

This section outlines the calibration procedure for the model parameters, describes the FX spot limit order book dataset, and presents empirical stylized facts used to motivate our modeling choices. We then report numerical experiments evaluating the resulting fill probability estimates at different price levels. Throughout this section, we adopt Model~III introduced in Section~\ref{examplesofmodels}, which is supported by the empirical findings presented below.

\subsection{Parameter Estimation}\label{sec: param_est}

The empirical analysis in Section~\ref{sec:data_analysis} indicates that both arrival and cancellation rates depend strongly on the prevailing spread size $S$. 
Following the estimation approach of Cont et al.\ (2010), we estimate the model parameters by sample averages over periods in which the spread remains equal to $S$.

\paragraph{Limit order arrivals.}
For limit orders posted at distance $\delta$ (in ticks) from the opposite best quote, conditional on $S$, we estimate the arrival rate by 
\begin{equation}\label{lambda_est}
    \hat{\lambda}^S(\delta)
    =
    \frac{N_l^S(\delta)}{T_*^S},
\end{equation}
where $N_l^S(\delta)$ denotes the total number of limit order arrivals observed at distance $\delta$ while the spread equals $S$, and $T_*^S$ is the total time (in seconds) during which the spread equals $S$ in the considered sample.
Under the symmetry assumption, the bid-side and ask-side rates are taken to be equal and given by $\tfrac{1}{2}\hat{\lambda}^S(\delta)$.

\paragraph{Market order arrivals.}
Similarly, the market order arrival rate conditional on spread size $S$ is estimated as
\begin{equation}\label{mu_est}
    \hat{\mu}^S
    =
    \frac{N_m^S}{T_*^S}\cdot \frac{S_m}{S_l},
\end{equation}
where $N_m^S$ is the total number of market orders observed while the spread equals $S$, and $S_m$ and $S_l$ denote the average sizes of market and limit orders, respectively. 
The ratio $\tfrac{S_m}{S_l}$ adjusts for the unit-size assumption in the model. 
Symmetry implies that the buy and sell market order intensities are both given by $\tfrac{1}{2}\hat{\mu}^S$. 
The empirical ratios $\tfrac{S_m}{S_l}$ are given in Table~\ref{tab: mark_lim_ratio}.

\begin{table}[h]
    \centering
    \begin{tabular}{r c c c c}\toprule
        & Week 1 & Week 2 & Week 3 & Week 4\\ \midrule
        $S_m / S_l$ & 0.40 & 0.47 & 0.38 & 0.43 \\ \bottomrule
    \end{tabular}
    \caption{Ratio of average market order size to average limit order size.}
    \label{tab: mark_lim_ratio}
\end{table}

\paragraph{Cancellation rates.}
Finally, we estimate the cancellation rate function $\theta(\delta,S)$ by
\begin{equation}\label{theta_est}
    \hat{\theta}^S(\delta)
    =
    \frac{N_c^S(\delta)}{T_*^S\,Q_\delta^S}\cdot \frac{S_c}{S_l},
\end{equation}
where $N_c^S(\delta)$ denotes the number of cancellations observed at distance $\delta$ from the opposite best quote while the spread equals $S$, and $S_c$ is the average cancellation size.
The normalization factor $Q_\delta^S$ corresponds to the average number of outstanding limit orders at that level, computed as
\[
Q_\delta^S
=
\frac{1}{2}\Bigl(Q_\delta^{S,\text{Bid}}+Q_\delta^{S,\text{Ask}}\Bigr),
\]
with
\begin{equation}
    Q_\delta^{S,\text{Bid}}
    =
    \frac{1}{S_l}\frac{1}{M}\sum_{j=1}^M V_{\delta}^{S,\text{Bid}}(j)
    \quad\text{and}\quad
    Q_\delta^{S,\text{Ask}}
    =
    \frac{1}{S_l}\frac{1}{M}\sum_{j=1}^M V_{\delta}^{S,\text{Ask}}(j),
\end{equation}
where $M$ denotes the number of observations in the sample, and $V_{\delta}^{S,\text{Bid}}(j)$ (resp.\ $V_{\delta}^{S,\text{Ask}}(j)$) is the total outstanding bid (resp.\ ask) volume at distance $\delta$ in the $j$th snapshot, conditional on spread $S$.
The ratio $\tfrac{S_c}{S_l}$ is given in Table~\ref{tab: can_lim_ratio}.

\begin{table}[h]
    \centering
    \begin{tabular}{r c c c c}\toprule
        & Week 1 & Week 2 & Week 3 & Week 4\\ \midrule
        $S_c / S_l$ & 1.00 & 1.00 & 1.00 & 1.00 \\ \bottomrule
    \end{tabular}
    \caption{Ratio of average cancellation size to average limit order size.}
    \label{tab: can_lim_ratio}
\end{table}

\subsection{Empirical Analysis of the FX Spot Market}\label{sec:data_analysis}

We conduct empirical analyses of the FX spot limit order book to motivate our intensity specifications and modeling assumptions. In particular, we study the empirical distribution of the bid--ask spread, assess the degree of symmetry between bid- and ask-side order flows, and quantify the relative frequency of submissions and cancellations.

Our dataset consists of high-frequency limit order book data from the trading venue LMAX, covering the period from 2--11--2020 to 29--10--2021. We focus on the EUR/USD currency pair, one of the most actively traded instruments in the global FX market.

Across four representative weeks (7--6--2021 to 2--7--2021), we find that the bid--ask spread lies predominantly between one and five ticks. In particular, the spread equals three or four ticks for approximately $80\%$ of the observed time, indicating that moderate spread sizes dominate typical market conditions.

Empirical estimates of limit order, market order, and cancellation intensities are given in Section~\ref{sec: param_est}.

\subsubsection{Order Flow Symmetry}

Model~III assumes symmetric order flow, meaning that buy and sell order arrival rates are comparable. To assess this assumption, we estimate empirical market order arrival rates on both sides of the book. Figure~\ref{fig: market_arr_tot} shows that buy and sell market order rates are consistently similar across all weeks, supporting the symmetry assumption and simplifying the calibration procedure.

\begin{figure}[h]
        \centering
        \begin{tikzpicture}
    \begin{axis}[
        ybar,
        ylabel={Arrival rate per second},
        ymin=-0.0005,
        width=10cm,
        height=4.5cm,
        bar width=0.4cm, 
        xtick=data,
        xticklabels={Week 1, Week 2, Week 3, Week 4},
        enlarge x limits=0.2,
        legend style={at={(0.5,0.92)}, anchor=north east,legend columns=-1},
        yticklabels={0.04, 0.08, 0.12, 0.16, 0.18},
        ytick={0.04, 0.08, 0.12, 0.16, 0.18},
        ymajorgrids=true, 
        extra y ticks={0}, 
        extra y tick style={grid=none},
        ]
        
        \addplot [fill=BarPurple, draw=BarPurple]  coordinates {
          (0, 0.105)          (1, 0.100)          (2, 0.118)    (3, 0.16)
        
        };
        \addplot [fill=BarOrange, color=BarOrange] coordinates {
          (0, 0.055) (1, 0.051)  (2, 0.059) (3, 0.084) 
        };
        
        \addplot [fill=BarBlue, color=BarBlue]  coordinates {
          (0, 0.05)          (1, 0.049)          (2, 0.059)   (3, 0.077)     
        };

        \legend{Total, Buy, Sell}
    
    \end{axis}
    \end{tikzpicture}
    \caption{Market order arrival rates (buy and sell) per second, by week.}
\label{fig: market_arr_tot}
\end{figure}

\subsubsection{Empirical Limit Order Executions}\label{sec: empirical_execution}

Figure~\ref{fig:execution_position} shows the empirical distribution of the distance between a limit order and the best quote at the time of execution. For buy (sell) orders, the distance is measured relative to the best bid (best ask). A distance of $0$ corresponds to executions at the best quote, while a strictly positive distance indicates executions caused by market orders consuming liquidity across multiple price levels. Across all weeks in the dataset, more than $90\%$ of executions occur at the best quotes, supporting the modeling assumption that market orders predominantly interact with the best bid and ask queues.

We next study the placement depth of executed limit orders at submission. Figure~\ref{fig:execution_position_start} shows the distribution of submission distances from the best quote, where negative values correspond to orders submitted inside the spread. We find that approximately $85\%$ of executed limit orders are submitted within one tick of the best quote, indicating that deeper placements contribute only marginally in practice.

\begin{figure}[h]
    \centering
    \begin{tikzpicture}
    \begin{axis}[
        ybar,
        xlabel={Distance in ticks from the best quote},
        ylabel={Percentage},
        height=4.5cm,
        ymin=-0.01,
        ymax=1.0,
        bar width=0.3cm, 
        xtick=data,
        xticklabels={0, 1, 2, 3, $>3$},
        enlarge x limits=0.15,
        legend style={at={(0.95,0.95)}, anchor=north east,legend columns=1},
        yticklabels={0.2, 0.4, 0.6, 0.8, 1.0},
        ytick={0.2, 0.4, 0.6, 0.8, 1.0},
        ymajorgrids=true, 
        extra y ticks={0}, 
        extra y tick style={grid=none},
        width=13cm,
        ]
        \addplot+ [fill=BarGreen, color=BarGreen]  coordinates {
          (0, 0.92)          (1, 0.05)          (2, 0.02)    (3, 0.01)   (4, 0.01)
        };
            
       \addplot+ [fill=BarOrange, color=BarOrange]  coordinates {
          (0, 0.90)          (1, 0.06)          (2, 0.02)    (3, 0.01)   (4, 0.01)
        };

        \addplot+ [fill=BarBlue, color=BarBlue]  coordinates {
          (0, 0.93)          (1, 0.04)          (2, 0.01)    (3, 0.01)   (4, 0.01)    
        };

        \addplot+ [fill=BarPurple, color=BarPurple]  coordinates {
          (0, 0.94)          (1, 0.03)          (2, 0.01)    (3, 0.01)   (4, 0.01)
        };
        \legend{Week 1, Week 2, Week 3, Week 4}
    
    \end{axis}
    \end{tikzpicture}
    \caption[Distance to best quote at execution for executed limit orders.]{Distribution of executed limit orders by their distance (in ticks) from the best quote at execution.}
    \label{fig:execution_position}
\end{figure}

\begin{figure}[H]
    \centering
    \begin{tikzpicture}
    \begin{axis}[
        ybar,
        xlabel={Distance in ticks from the best quote},
        ylabel={Percentage},
        height=4.5cm,
        ymin=-0.01,
        ymax=0.6,
        bar width=0.3cm, 
        xtick=data,
        xticklabels={$<0$, 0, 1, 2, 3, $>3$},
        enlarge x limits=0.15,
        legend style={at={(0.95,0.95)}, anchor=north east,legend columns=1},
        yticklabels={0.1, 0.2, 0.3, 0.4, 0.5, 0.6},
        ytick={0.1, 0.2, 0.3, 0.4, 0.5, 0.6},
        ymajorgrids=true, 
        extra y ticks={0}, 
        extra y tick style={grid=none},
        width=13cm,
        ]
        \addplot+ [fill=BarGreen, color=BarGreen]  coordinates {
          (0, 0.42)          (1, 0.27)          (2, 0.17)    (3, 0.08)   (4, 0.03)  (5, 0.02)
        };
            
       \addplot+ [fill=BarOrange, color=BarOrange]  coordinates {
          (0, 0.38)          (1, 0.30)          (2, 0.17)    (3, 0.09)   (4, 0.03)  (5, 0.02)
        };

        \addplot+ [fill=BarBlue, color=BarBlue]  coordinates {
          (0, 0.46)          (1, 0.28)          (2, 0.16)    (3, 0.06)   (4, 0.02)  (5, 0.02)
        };

        \addplot+ [fill=BarPurple, color=BarPurple]  coordinates {
          (0, 0.56)          (1, 0.23)          (2, 0.12)    (3, 0.06)   (4, 0.02)  (5, 0.01)
        };
        \legend{Week 1, Week 2, Week 3, Week 4}
    
    \end{axis}
    \end{tikzpicture}
    \caption[Distance to best quote at submission for executed limit orders.]{Distribution of executed limit orders by their distance (in ticks) from the best quote at submission.}

    \label{fig:execution_position_start}
\end{figure}

\subsubsection{Empirical Arrival and Cancellation Rates}\label{sec:emp_rates}

We next quantify order flow as a function of the spread size $S$ and the distance (in ticks) to the opposite best quote. Figure~\ref{fig:limit_rate_spread} shows empirical limit order arrival rates for $S\in\{1,\dots,5\}$, showing substantially higher activity when the spread is small. Figure~\ref{fig:marketarrivalrate} presents market order arrival rates by spread size, which similarly decrease rapidly as $S$ increases. Figure~\ref{fig:cancel_rate_spread} shows that cancellation intensities also depend strongly on $S$ and closely mirror the submission profiles, consistent with the fact that nearly all limit orders are eventually cancelled (approximately $99.9\%$; see Table~\ref{tab: cancel_percentage}). These findings motivate Model~III, in which all order flow intensities explicitly depend on $S$.

\begin{figure}[!htbp]
\centering
\begin{subfigure}{\textwidth}
    \centering
    \begin{tikzpicture}[thick,scale=0.9, every node/.style={scale=0.9}]
    \begin{axis}[
        ybar,
        xlabel={Distance in ticks from the opposite best quote},
        ylabel={Arrival rate per second},
        ymin=-0.1,
        ymax=80,
        width=17cm,
        height=4.5cm,
        bar width=0.05cm, 
        xtick=data,
        xticklabels={$1$, $2$, $3$, $4$, $5$, $6$, $7$, $8$, $9$, $10$, $11$, $12$, $13$, $14$, $15$},
        enlarge x limits=0.04,
        legend style={at={(0.99,0.95)}, anchor=north east,legend columns=1},
        ytick={20, 40, 60, 80},
        ymajorgrids=true, 
        extra y ticks={0},
        extra y tick style={grid=none}
        ]

        \addplot+ [fill=BarGreen, color=BarGreen] coordinates {
          (0,13.27) (1,34.72) (2,59.54) (3,66.5) (4,69.78) (5,50.61) (6,34.53) (7,23.19) (8,17.61) (9,16.93) (10,13.04) (11,9.25) (12,7.36) (13,5.91) (14,5.02)
        };
        
        \addplot+ [fill=BarOrange, color=BarOrange] coordinates {
          (0,1.27) (1,7.75) (2,20.22) (3,26.18) (4,24.48) (5,22.09) (6,15.53) (7,10.9) (8,7.14) (9,6.6) (10,5.19) (11,3.92) (12,3.19) (13,2.72) (14,2.23)
        };
        
        \addplot+ [fill=BarBlue, color=BarBlue] coordinates {
          (0,0.08) (1,1.09) (2,5.73) (3,11.6) (4,13.77) (5,11.44) (6,8.15) (7,4.94) (8,2.94) (9,2.35) (10,2.4) (11,1.72) (12,1.66) (13,1.4) (14,1.18)
        };
        
        \addplot+ [fill=BarPurple, color=BarPurple] coordinates {
          (0,0.02) (1,0.11) (2,2.27) (3,7.17) (4,10.54) (5,10.43) (6,7.52) (7,5.3) (8,3.61) (9,2.04) (10,2.03) (11,1.4) (12,1.43) (13,1.22) (14,1.09)
        };
        
        \addplot+ [fill=BarYellow, color=BarYellow] coordinates {
          (0,0.02) (1,0.05) (2,0.3) (3,5.48) (4,11.84) (5,13.96) (6,12.57) (7,9.62) (8,7.47) (9,5.89) (10,3.42) (11,3.24) (12,2.34) (13,2.26) (14,1.87)
        };        
        
        \legend{$S=1$, $S=2$, $S=3$, $S=4$, $S=5$}

    \end{axis}
    \end{tikzpicture}
    \caption{Rates between 7--6--2021 and 11--6--2021}
\end{subfigure}
\\
\begin{subfigure}{\textwidth}
    \centering
    \begin{tikzpicture}[thick,scale=0.9, every node/.style={scale=0.9}]
    \begin{axis}[
        ybar,
        xlabel={Distance in ticks from the opposite best quote},
        ylabel={Arrival rate per second},
        ymin=-0.1,
        ymax=200,
        width=17cm,
        height=4.5cm,
        bar width=0.05cm, 
        xtick=data,
        xticklabels={$0$, $1$, $2$, $3$, $4$, $5$, $6$, $7$, $8$, $9$, $10$, $11$, $12$, $13$, $14$, $15$},
        enlarge x limits=0.04,
        legend style={at={(0.99,0.95)}, anchor=north east,legend columns=1},
        ytick={50, 100, 150, 200},
        ymajorgrids=true, 
        extra y ticks={0},
        extra y tick style={grid=none}
        ]
        
        \addplot+ [fill=BarGreen, color=BarGreen] coordinates {
          (0,32.88) (1,73.2) (2,144.32) (3,169.15) (4,193.8) (5,139.58) (6,107.18) (7,61.2) (8,52.17) (9,44.57) (10,33.57) (11,22.95) (12,17.92) (13,14.34) (14,12.54)
        };
        
        \addplot+ [fill=BarOrange, color=BarOrange] coordinates {
          (0,2.45) (1,14.79) (2,38.31) (3,57.27) (4,55.74) (5,57.31) (6,39.16) (7,29.01) (8,16.88) (9,16.51) (10,12.14) (11,8.73) (12,6.71) (13,5.82) (14,5.13)
        };
        
        \addplot+ [fill=BarBlue, color=BarBlue] coordinates {
          (0,0.12) (1,1.43) (2,8.51) (3,19.31) (4,23.6) (5,22.97) (6,16.94) (7,11.53) (8,5.67) (9,5.52) (10,4.95) (11,3.29) (12,2.83) (13,2.38) (14,2.24)
        };
        
        \addplot+ [fill=BarPurple, color=BarPurple] coordinates {
          (0,0.02) (1,0.1) (2,1.88) (3,7.97) (4,13.06) (5,13.47) (6,11.28) (7,7.82) (8,5.75) (9,3.08) (10,3.06) (11,2.02) (12,1.66) (13,1.33) (14,1.24)
        };
        
        \addplot+ [fill=BarYellow, color=BarYellow] coordinates {
          (0,0.01) (1,0.04) (2,0.21) (3,5.56) (4,11.93) (5,15.14) (6,14.41) (7,10.96) (8,8.99) (9,6.97) (10,3.49) (11,2.85) (12,1.74) (13,1.5) (14,1.13)
        };
        
        \legend{$S=1$, $S=2$, $S=3$, $S=4$, $S=5$}
    \end{axis}
    \end{tikzpicture}
    \caption{Rates between 14--6--2021 and 18--6--2021}
\end{subfigure}
\\\begin{subfigure}{\textwidth}
    \centering
    \begin{tikzpicture}[thick,scale=0.9, every node/.style={scale=0.9}]
    \begin{axis}[
        ybar,
        xlabel={Distance in ticks from the opposite best quote},
        ylabel={Arrival rate per second},
        ymin=-0.1,
        ymax=40,
        width=17cm,
        height=4.5cm,
        bar width=0.05cm, 
        xtick=data,
        xticklabels={$0$, $1$, $2$, $3$, $4$, $5$, $6$, $7$, $8$, $9$, $10$, $11$, $12$, $13$, $14$, $15$},
        enlarge x limits=0.04,
        legend style={at={(0.99,0.95)}, anchor=north east,legend columns=1},
        ytick={10, 20, 30, 40},
        ymajorgrids=true, 
        extra y ticks={0},
        extra y tick style={grid=none}
        ]
        
        \addplot+ [fill=BarGreen, color=BarGreen] coordinates {
          (0,6.6) (1,16.07) (2,29.68) (3,34.59) (4,34.96) (5,26.07) (6,19.36) (7,11.87) (8,11.52) (9,8.3) (10,5.95) (11,3.98) (12,2.77) (13,2.22) (14,1.98)
        };
        
        \addplot+ [fill=BarOrange, color=BarOrange] coordinates {
          (0,1.74) (1,9.32) (2,24.18) (3,38.28) (4,36.22) (5,32.48) (6,22.0) (7,15.04) (8,11.95) (9,10.15) (10,6.8) (11,4.61) (12,3.47) (13,2.68) (14,2.18)
        };
        
        \addplot+ [fill=BarBlue, color=BarBlue] coordinates {
          (0,0.1) (1,1.44) (2,7.64) (3,18.51) (4,22.6) (5,20.79) (6,14.6) (7,8.95) (8,5.46) (9,5.27) (10,3.78) (11,2.35) (12,1.99) (13,1.57) (14,1.41)
        };
        
        \addplot+ [fill=BarPurple, color=BarPurple] coordinates {
         (0,0.02) (1,0.1) (2,2.37) (3,8.17) (4,15.0) (5,15.14) (6,12.36) (7,7.57) (8,5.23) (9,3.79) (10,2.89) (11,1.65) (12,1.36) (13,1.11) (14,1.04)
        };
        
        \addplot+ [fill=BarYellow, color=BarYellow] coordinates {
          (0,0.01) (1,0.03) (2,0.16) (3,4.88) (4,9.88) (5,12.59) (6,12.26) (7,8.36) (8,5.82) (9,5.23) (10,3.15) (11,1.96) (12,1.13) (13,1.04) (14,0.87)
        };
        
        \legend{$S=1$, $S=2$, $S=3$, $S=4$, $S=5$}
    \end{axis}
    \end{tikzpicture}
    \caption{Rates between 21--6--2021 and 25--6--2021}
\end{subfigure}
\\\begin{subfigure}{\textwidth}
    \centering
    \begin{tikzpicture}[thick,scale=0.9, every node/.style={scale=0.9}]
    \begin{axis}[
        ybar,
        xlabel={Distance in ticks from the opposite best quote},
        ylabel={Arrival rate per second},
        ymin=-0.1,
        ymax=30,
        width=17cm,
        height=4.5cm,
        bar width=0.05cm, 
        xtick=data,
        xticklabels={$0$, $1$, $2$, $3$, $4$, $5$, $6$, $7$, $8$, $9$, $10$, $11$, $12$, $13$, $14$, $15$},
        enlarge x limits=0.04,
        legend style={at={(0.99,0.95)}, anchor=north east,legend columns=1},
        ytick={7.5, 15, 22.5, 30},
        ymajorgrids=true, 
        extra y ticks={0},
        extra y tick style={grid=none}
        ]
        
        \addplot+ [fill=BarGreen, color=BarGreen] coordinates {
          (0,5.64) (1,12.94) (2,21.38) (3,23.3) (4,24.85) (5,18.64) (6,15.3) (7,8.59) (8,8.78) (9,6.04) (10,4.3) (11,3.0) (12,1.98) (13,1.47) (14,1.21)
        };
        
        \addplot+ [fill=BarOrange, color=BarOrange] coordinates {
         (0,1.31) (1,5.36) (2,11.26) (3,17.21) (4,16.07) (5,13.55) (6,10.03) (7,7.34) (8,4.89) (9,3.86) (10,2.84) (11,2.0) (12,1.49) (13,1.01) (14,0.81)
        };
        
        \addplot+ [fill=BarBlue, color=BarBlue] coordinates {
         (0,0.2) (1,2.21) (2,6.95) (3,13.62) (4,17.02) (5,16.99) (6,10.19) (7,8.05) (8,4.54) (9,3.3) (10,2.91) (11,1.86) (12,1.62) (13,1.16) (14,0.95)
        };
        
        \addplot+ [fill=BarPurple, color=BarPurple] coordinates {
          (0,0.08) (1,0.14) (2,3.79) (3,9.02) (4,15.12) (5,16.39) (6,12.82) (7,8.66) (8,6.94) (9,3.27) (10,2.91) (11,1.93) (12,1.79) (13,1.34) (14,1.05)
        };
        
        \addplot+ [fill=BarYellow, color=BarYellow] coordinates {
          (0,0.04) (1,0.05) (2,0.25) (3,6.5) (4,11.7) (5,14.46) (6,14.05) (7,9.62) (8,8.02) (9,6.23) (10,2.89) (11,2.32) (12,1.85) (13,1.8) (14,1.17)
        };
        
        \legend{$S=1$, $S=2$, $S=3$, $S=4$, $S=5$}
    \end{axis}
    \end{tikzpicture}
    \caption{Rates between 28--6--2021 and 2--7--2021}
\end{subfigure}
\caption{Limit order arrival rates by distance $\delta$ (in ticks) from the opposite best quote, for each spread size $S$.}
\label{fig:limit_rate_spread}
\end{figure}

\begin{figure}[!htbp]
    \centering
    \begin{tikzpicture}
    \begin{axis}[
        ybar,
        xlabel={Spread size (in ticks)},
        ylabel={Arrival rate per second},
        height=4.5cm,
        ymin=-0.01,
        ymax=6.5,
        bar width=0.3cm, 
        xtick=data,
        xticklabels={1, 2, 3, 4, 5},
        enlarge x limits=0.15,
        legend style={at={(0.95,0.95)}, anchor=north east,legend columns=1},
        yticklabels={1.0, 2.0, 3.0, 4.0, 5.0, 6.0},
        ytick={1.0, 2.0, 3.0, 4.0, 5.0, 6.0},
        ymajorgrids=true, 
        extra y ticks={0}, 
        extra y tick style={grid=none},
        width=13cm,
        ]
        \addplot+ [fill=BarGreen, color=BarGreen]  coordinates {
          (0, 2.58)          (1, 0.35)          (2, 0.08)    (3, 0.05)   (4, 0.08)    
        };
            
       \addplot+ [fill=BarOrange, color=BarOrange]  coordinates {
          (0, 5.99)          (1, 0.81)          (2, 0.13)    (3, 0.06)   (4, 0.06)    
        };

        \addplot+ [fill=BarBlue, color=BarBlue]  coordinates {
          (0, 1.26)          (1, 0.52)          (2, 0.14)    (3, 0.08)   (4, 0.07)    
        };

        \addplot+ [fill=BarPurple, color=BarPurple]  coordinates {
          (0, 0.55)          (1, 0.19)          (2, 0.12)    (3, 0.1)   (4, 0.11)    
        };
        \legend{Week 1, Week 2, Week 3, Week 4}
    
    \end{axis}
    \end{tikzpicture}
    \caption[Market order arrival rates.]{Market order arrival rates (per second) for spread sizes $S=1,\dots,5$ ticks.}
    \label{fig:marketarrivalrate}
\end{figure}

\begin{figure}[!htbp]
\centering
\begin{subfigure}{\textwidth}
    \centering
    \begin{tikzpicture}[thick,scale=0.9, every node/.style={scale=0.9}]
    \begin{axis}[
        ybar,
        xlabel={Distance in ticks from the opposite best quote},
        ylabel={Cancellation rate per second},
        ymin=-0.1,
        ymax=70,
        width=17cm,
        height=4.5cm,
        bar width=0.05cm, 
        xtick=data,
        xticklabels={$1$, $2$, $3$, $4$, $5$, $6$, $7$, $8$, $9$, $10$, $11$, $12$, $13$, $14$, $15$},
        enlarge x limits=0.04,
        legend style={at={(0.99,0.95)}, anchor=north east,legend columns=1},
        ytick={0, 20, 40, 60},
        ymajorgrids=true, 
        extra y ticks={0}, 
        extra y tick style={grid=none},
        ]
        
        \addplot+ [fill=BarGreen, color=BarGreen] coordinates {
          (0,43.95) (1,41.11) (2,53.86) (3,60.57) (4,55.1) (5,52.11) (6,41.01) (7,31.82) (8,21.99) (9,14.99) (10,13.33) (11,10.39) (12,7.98) (13,6.72) (14,5.97)
        };
        
        \addplot+ [fill=BarOrange, color=BarOrange] coordinates {
          (0,0) (1,18.77) (2,19.63) (3,25.98) (4,25.68) (5,20.98) (6,16.12) (7,9.99) (8,7.4) (9,5.65) (10,5.08) (11,4.05) (12,3.47) (13,2.87) (14,2.35)
        };
        
        \addplot+ [fill=BarBlue, color=BarBlue] coordinates {
         (0,0) (1,0) (2,9.25) (3,11.27) (4,12.65) (5,10.22) (6,7.21) (7,5.79) (8,4.35) (9,3.06) (10,2.14) (11,1.87) (12,1.61) (13,1.49) (14,1.17)
        };
        
        \addplot+ [fill=BarPurple, color=BarPurple] coordinates {
          (0,0) (1,0) (2,0) (3,7.91) (4,10.76) (5,10.46) (6,7.5) (7,4.88) (8,3.2) (9,2.4) (10,1.69) (11,1.62) (12,1.3) (13,1.29) (14,1.08)
        };
        
        \addplot+ [fill=BarYellow, color=BarYellow] coordinates {
          (0,0) (1,0) (2,0) (3,0) (4,12.72) (5,14.92) (6,13.44) (7,9.69) (8,6.51) (9,4.71) (10,3.74) (11,2.63) (12,2.46) (13,2.16) (14,1.91)
        };
        
        \legend{$S=1$, $S=2$, $S=3$, $S=4$, $S=5$}
    \end{axis}
    \end{tikzpicture}
    \caption{Rates between 7--6--2021 and 11--6--2021.}
\end{subfigure}
\\
\begin{subfigure}{\textwidth}
    \centering
    \begin{tikzpicture}[thick,scale=0.9, every node/.style={scale=0.9}]
    \begin{axis}[
        ybar,
        xlabel={Distance in ticks from the opposite best quote},
        ylabel={Cancellation rate per second},
        ymin=-0.1,
        ymax=160,
        width=17cm,
        height=4.5cm,
        bar width=0.05cm, 
        xtick=data,
        xticklabels={$1$, $2$, $3$, $4$, $5$, $6$, $7$, $8$, $9$, $10$, $11$, $12$, $13$, $14$, $15$},
        enlarge x limits=0.04,
        legend style={at={(0.99,0.95)}, anchor=north east,legend columns=1},
        ytick={25, 50, 75, 100, 125, 150},
        ymajorgrids=true, 
        extra y ticks={0}, 
        extra y tick style={grid=none},
        ]
        
        \addplot+ [fill=BarGreen, color=BarGreen] coordinates {
          (0,107.95) (1,99.02) (2,125.4) (3,153.33) (4,141.69) (5,140.91) (6,123.33) (7,92.04) (8,59.04) (9,40.4) (10,34.88) (11,26.99) (12,20.54) (13,17.72) (14,14.99)
        };
        
        \addplot+ [fill=BarOrange, color=BarOrange] coordinates {
          (0,0) (1,40.24) (2,39.18) (3,53.06) (4,60.38) (5,46.46) (6,43.74) (7,27.51) (8,20.3) (9,13.41) (10,11.69) (11,9.56) (12,7.29) (13,6.5) (14,5.7)
        };
        
        \addplot+ [fill=BarBlue, color=BarBlue] coordinates {
         (0,0) (1,0) (2,17.72) (3,18.78) (4,21.79) (5,18.72) (6,14.91) (7,13.4) (8,10.06) (9,5.87) (10,4.58) (11,3.75) (12,2.76) (13,2.59) (14,2.31)
        };
        
        \addplot+ [fill=BarPurple, color=BarPurple] coordinates {
         (0,0) (1,0) (2,0) (3,9.96) (4,13.01) (5,13.57) (6,10.2) (7,7.78) (8,5.26) (9,4.1) (10,2.67) (11,2.32) (12,1.55) (13,1.39) (14,1.25)
        };
        
        \addplot+ [fill=BarYellow, color=BarYellow] coordinates {
          (0,0) (1,0) (2,0) (3,0) (4,12.23) (5,15.5) (6,16.15) (7,10.96) (8,9.06) (9,4.41) (10,3.73) (11,2.58) (12,1.92) (13,1.33) (14,1.23)
        };
        
        \legend{$S=1$, $S=2$, $S=3$, $S=4$, $S=5$}
    \end{axis}
    \end{tikzpicture}
    \caption{Rates between 14--6--2021 and 18--6--2021.}
\end{subfigure}
\\\begin{subfigure}{\textwidth}
    \centering
    \begin{tikzpicture}[thick,scale=0.9, every node/.style={scale=0.9}]
    \begin{axis}[
        ybar,
        xlabel={Distance in ticks from the opposite best quote},
        ylabel={Cancellation rate per second},
        ymin=-0.1,
        ymax=40,
        width=17cm,
        height=4.5cm,
        bar width=0.05cm, 
        xtick=data,
        xticklabels={$1$, $2$, $3$, $4$, $5$, $6$, $7$, $8$, $9$, $10$, $11$, $12$, $13$, $14$, $15$},
        enlarge x limits=0.04,
        legend style={at={(0.99,0.95)}, anchor=north east,legend columns=1},
        ytick={5,10,15, 20, 25, 30, 35, 40},
        ymajorgrids=true, 
        extra y ticks={0}, 
        extra y tick style={grid=none},
        ]
        
        \addplot+ [fill=BarGreen, color=BarGreen] coordinates {
          (0,19.41) (1,20.44) (2,26.32) (3,30.5) (4,29.63) (5,26.57) (6,21.48) (7,16.25) (8,10.89) (9,8.04) (10,5.96) (11,4.58) (12,3.17) (13,2.67) (14,2.35)
        };
        
        \addplot+ [fill=BarOrange, color=BarOrange] coordinates {
          (0,0) (1,24.63) (2,27.37) (3,33.8) (4,37.12) (5,29.62) (6,24.74) (7,16.36) (8,11.41) (9,8.27) (10,6.53) (11,5.08) (12,3.64) (13,3.08) (14,2.57)
        };
        
        \addplot+ [fill=BarBlue, color=BarBlue] coordinates {
          (0,0) (1,0) (2,15.7) (3,18.26) (4,20.5) (5,18.26) (6,13.29) (7,10.48) (8,8.42) (9,5.15) (10,3.6) (11,2.66) (12,1.91) (13,1.77) (14,1.47)
        };
        
        \addplot+ [fill=BarPurple, color=BarPurple] coordinates {
         (0,0) (1,0) (2,0) (3,11.47) (4,14.19) (5,14.81) (6,11.28) (7,7.88) (8,5.19) (9,3.98) (10,2.81) (11,1.96) (12,1.26) (13,1.17) (14,1.07)
        };
        
        \addplot+ [fill=BarYellow, color=BarYellow] coordinates {
          (0,0) (1,0) (2,0) (3,0) (4,10.06) (5,12.89) (6,12.47) (7,9.22) (8,5.93) (9,3.71) (10,2.9) (11,1.91) (12,1.26) (13,0.92) (14,0.94)
        };
        
        \legend{$S=1$, $S=2$, $S=3$, $S=4$, $S=5$}
    \end{axis}
    \end{tikzpicture}
    \caption{Rates between 21--6--2021 and 25--6--2021.}
\end{subfigure}
\\\begin{subfigure}{\textwidth}
    \centering
    \begin{tikzpicture}[thick,scale=0.9, every node/.style={scale=0.9}]
    \begin{axis}[
        ybar,
        xlabel={Distance in ticks from the opposite best quote},
        ylabel={Cancellation rate per second},
        ymin=-0.1,
        ymax=20,
        width=17cm,
        height=4.5cm,
        bar width=0.05cm, 
        xtick=data,
        xticklabels={$1$, $2$, $3$, $4$, $5$, $6$, $7$, $8$, $9$, $10$, $11$, $12$, $13$, $14$, $15$},
        enlarge x limits=0.04,
        legend style={at={(0.99,0.95)}, anchor=north east,legend columns=1},
        ytick={2.5, 5, 7.5, 10, 12.5, 15, 17.5},
        ymajorgrids=true, 
        extra y ticks={0}, 
        extra y tick style={grid=none},
        ]
        
        \addplot+ [fill=BarGreen, color=BarGreen] coordinates {
          (0, 7.70) (1, 9.26) (2, 12.64) (3, 14.10) (4, 13.14) (5, 10.84) (6, 9.65) (7, 7.90) (8, 4.61) (9, 3.81) (10, 2.74) (11, 1.99) (12, 1.41) (13, 1.03) (14, 0.82)
        };
        
        \addplot+ [fill=BarOrange, color=BarOrange] coordinates {
         (0,0) (1,8.97) (2,13.03) (3,16.08) (4,18.42) (5,12.63) (6,10.89) (7,6.31) (8,4.19) (9,3.76) (10,2.62) (11,1.98) (12,1.43) (13,1.12) (14,0.86)
        };
        
        \addplot+ [fill=BarBlue, color=BarBlue] coordinates {
         (0,0) (1,0) (2,10.12) (3,14.11) (4,16.21) (5,14.11) (6,10.44) (7,8.63) (8,6.23) (9,3.67) (10,2.66) (11,2.14) (12,1.57) (13,1.39) (14,0.99)
        };
        
        \addplot+ [fill=BarPurple, color=BarPurple] coordinates {
          (0,0) (1,0) (2,0) (3,11.45) (4,14.72) (5,16.47) (6,12.26) (7,9.11) (8,5.34) (9,4.36) (10,2.59) (11,2.16) (12,1.58) (13,1.4) (14,1.14)
        };
        
        \addplot+ [fill=BarYellow, color=BarYellow] coordinates {
          (0,0) (1,0) (2,0) (3,0) (4,11.57) (5,14.84) (6,15.32) (7,10.4) (8,7.37) (9,4.14) (10,3.47) (11,2.16) (12,1.96) (13,1.32) (14,1.24)
        };
        
        \legend{$S=1$, $S=2$, $S=3$, $S=4$, $S=5$}
    \end{axis}
    \end{tikzpicture}
    \caption{Rates between 28--6--2021 and 2--7--2021.}
\end{subfigure}
\caption{Cancellation rates by distance $\delta$ (ticks) from the opposite best quote, for each spread size $S$.}
  \label{fig:cancel_rate_spread}
\end{figure}

\begin{table}[h]
    \centering
        \begin{tabular}{r r r r r}\toprule
            & Week 1 & Week 2 & Week 3 & Week 4\\ \midrule
         Cancelled & 99.91\% & 99.93\% & 99.91\% & 99.87\% \\
         (Partially) filled & 0.09\% & 0.07\% & 0.09\% & 0.13\% \\ \bottomrule
       \end{tabular}
  
   \caption{Percentage of limit orders that are cancelled or (partially) filled.}
    \label{tab: cancel_percentage}
\end{table}

\subsubsection{Average Number of Outstanding Orders}\label{sec:avg_outstanding}

Figure~\ref{fig:average_q} shows the time-averaged number of outstanding limit orders at price levels located one to ten ticks from the opposite best quote. The queue profiles are highly stable across the four sampled weeks. For each spread size, the best quote has the smallest average queue size, which increases with depth and peaks around four to six ticks from the best quote. Beyond this point, average queue sizes decline and remain approximately constant at deeper levels.

\begin{figure}[!htbp]
\centering
\begin{subfigure}[b]{0.49\textwidth}
    \centering
    \begin{tikzpicture}
    \begin{axis}[
        ybar,
        xlabel={\small Distance in ticks from the opposite best quote},
        ylabel={Average quantity outstanding},
        y label style={at={(axis description cs:0.13,0.5)}, anchor=south},
        ymin=-20000,
        ymax=4500000,
        bar width=0.04cm, 
        xtick=data,
        xticklabels={$1$, $2$, $3$, $4$, $5$, $6$, $7$, $8$, $9$, $10$},
        enlarge x limits=0.07,
        legend style={at={(0.02,0.98)}, anchor=north west,legend columns=1},
        ytick={1000000,2000000,3000000,4000000},
        ymajorgrids=true, 
        extra y ticks={0}, 
        extra y tick style={grid=none},
        major tick length = 0.08cm
        ]
        
        \addplot+ [fill=BarGreen, color=BarGreen] coordinates {
         (0, 881009) (1, 1276706) (2, 2211293) (3, 2896346) (4, 3217561) (5, 2929171) (6, 2517662) (7, 2308628) (8, 2249335) (9, 2167686)
        };
        
        \addplot+ [fill=BarOrange, color=BarOrange] coordinates {
          (0, 0) (1, 987704) (2, 1915617) (3, 2954285) (4, 3395169) (5, 3358261) (6, 2852408) (7, 2303484) (8, 2295585) (9, 2248940)
        };
        
        \addplot+ [fill=BarBlue, color=BarBlue] coordinates {
         (0, 0) (1, 0) (2, 1320495) (3, 2720061) (4, 3596108) (5, 3786409) (6, 3222420) (7, 2466837) (8, 2122430) (9, 2223011)
        };
        
        \addplot+ [fill=BarPurple, color=BarPurple] coordinates {
          (0, 0) (1, 0) (2, 0) (3, 2105311) (4, 3215664) (5, 3759231) (6, 3567034) (7, 2909708) (8, 2393342) (9, 2111617)
        };
        
        \addplot+ [fill=BarYellow, color=BarYellow] coordinates {
          (0, 0) (1, 0) (2, 0) (3, 0) (4, 2668738) (5, 3221010) (6, 3483012) (7, 3076086) (8, 2669657) (9, 2305588)
        };
        
        \legend{$S=1$, $S=2$, $S=3$, $S=4$, $S=5$}
    \end{axis}
    \end{tikzpicture}
    \caption{Average quantities between 7--6--2021 and \\11--6--2021.}
\end{subfigure}
\begin{subfigure}[b]{0.49\textwidth}
    \centering
    \begin{tikzpicture}
    \begin{axis}[
        ybar,
        xlabel={\small Distance in ticks from the opposite best quote},
        ylabel={Average quantity outstanding},
        y label style={at={(axis description cs:0.13,0.5)}, anchor=south},
        ymin=-20000,
        ymax=4500000,
        bar width=0.04cm, 
        xtick=data,
        xticklabels={$1$, $2$, $3$, $4$, $5$, $6$, $7$, $8$, $9$, $10$},
        enlarge x limits=0.07,
        legend style={at={(0.02,0.98)}, anchor=north west,legend columns=1},
        ytick={1000000,2000000,3000000,4000000},
        ymajorgrids=true, 
        extra y ticks={0}, 
        extra y tick style={grid=none},
        major tick length = 0.08cm
        ]
        
        \addplot+ [fill=BarGreen, color=BarGreen] coordinates {
          (0, 937201) (1, 1178273) (2, 1980725) (3, 2666660) (4, 3180881) (5, 2948543) (6, 2664931) (7, 2441223) (8, 2320286) (9, 2252619)
        };
        
        \addplot+ [fill=BarOrange, color=BarOrange] coordinates {
         (0, 0) (1, 1009250) (2, 1704020) (3, 2768598) (4, 3173786) (5, 3461039) (6, 2937258) (7, 2393703) (8, 2391245) (9, 2334668)
        };
        
        \addplot+ [fill=BarBlue, color=BarBlue] coordinates {
         (0, 0) (1, 0) (2, 1197928) (3, 2554048) (4, 3338846) (5, 3776215) (6, 3391919) (7, 2535485) (8, 2204979) (9, 2345263)
        };
        
        \addplot+ [fill=BarPurple, color=BarPurple] coordinates {
         (0, 0) (1, 0) (2, 0) (3, 1896755) (4, 3125428) (5, 3548816) (6, 3753704) (7, 3079049) (8, 2450451) (9, 2188514)
        };
        
        \addplot+ [fill=BarYellow, color=BarYellow] coordinates {
          (0, 0) (1, 0) (2, 0) (3, 0) (4, 2753656) (5, 3140941) (6, 3523565) (7, 3450564) (8, 2877668) (9, 2358425)
        };
        
        \legend{$S=1$, $S=2$, $S=3$, $S=4$, $S=5$}
    \end{axis}
    \end{tikzpicture}
    \caption{Average quantities between 14--6--2021 and \\18--6--2021.}
\end{subfigure}
\\\begin{subfigure}[b]{0.49\textwidth}
    \centering
    \begin{tikzpicture}
    \begin{axis}[
        ybar,
        xlabel={\small Distance in ticks from the opposite best quote},
        ylabel={Average quantity outstanding},
        y label style={at={(axis description cs:0.13,0.5)}, anchor=south},
        ymin=-20000,
        ymax=4500000,
        bar width=0.04cm, 
        xtick=data,
        xticklabels={$1$, $2$, $3$, $4$, $5$, $6$, $7$, $8$, $9$, $10$},
        enlarge x limits=0.07,
        legend style={at={(0.02,0.98)}, anchor=north west,legend columns=1},
        ytick={1000000,2000000,3000000,4000000},
        ymajorgrids=true, 
        extra y ticks={0}, 
        extra y tick style={grid=none},
        major tick length = 0.08cm
        ]
        
        \addplot+ [fill=BarGreen, color=BarGreen] coordinates {
          (0, 946203) (1, 1216197) (2, 2019834) (3, 2611302) (4, 3050296) (5, 2880843) (6, 2631826) (7, 2374942) (8, 2322966) (9, 2255550)
        };
        
        \addplot+ [fill=BarOrange, color=BarOrange] coordinates {
          (0, 0) (1, 997756) (2, 1719178) (3, 2794963) (4, 3157827) (5, 3280082) (6, 2792516) (7, 2341025) (8, 2363071) (9, 2339150)
        };
        
        \addplot+ [fill=BarBlue, color=BarBlue] coordinates {
          (0, 0) (1, 0) (2, 1171872) (3, 2613741) (4, 3363508) (5, 3637893) (6, 3205178) (7, 2435920) (8, 2196744) (9, 2334744)
        };
        
        \addplot+ [fill=BarPurple, color=BarPurple] coordinates {
         (0, 0) (1, 0) (2, 0) (3, 1854415) (4, 3172715) (5, 3551835) (6, 3582193) (7, 2887714) (8, 2356096) (9, 2218379)
        };
        
        \addplot+ [fill=BarYellow, color=BarYellow] coordinates {
          (0, 0) (1, 0) (2, 0) (3, 0) (4, 2702744) (5, 3184181) (6, 3524908) (7, 3297543) (8, 2756338) (9, 2309784)
        };
        
        \legend{$S=1$, $S=2$, $S=3$, $S=4$, $S=5$}
    \end{axis}
    \end{tikzpicture}
    \caption{Average quantities between 21--6--2021 and \\25--6--2021.}
\end{subfigure}
\begin{subfigure}[b]{0.49\textwidth}
    \centering
    \begin{tikzpicture}
    \begin{axis}[
        ybar,
        xlabel={\small Distance in ticks from the opposite best quote},
        ylabel={Average quantity outstanding},
        y label style={at={(axis description cs:0.13,0.5)}, anchor=south},
        ymin=-20000,
        ymax=4500000,
        bar width=0.04cm, 
        xtick=data,
        xticklabels={$1$, $2$, $3$, $4$, $5$, $6$, $7$, $8$, $9$, $10$},
        enlarge x limits=0.07,
        legend style={at={(0.02,0.98)}, anchor=north west,legend columns=1},
        ytick={1000000,2000000,3000000,4000000},
        ymajorgrids=true, 
        extra y ticks={0}, 
        extra y tick style={grid=none},
        major tick length = 0.08cm
        ]
        
        \addplot+ [fill=BarGreen, color=BarGreen] coordinates {
          (0, 934078) (1, 1314196) (2, 2159477) (3, 2748123) (4, 3227522) (5, 3055098) (6, 2559423) (7, 2289031) (8, 2384276) (9, 2310791)
        };
        
        \addplot+ [fill=BarOrange, color=BarOrange] coordinates {
         (0, 0) (1, 1026808) (2, 1875125) (3, 2831906) (4, 3238382) (5, 3495244) (6, 2892744) (7, 2329283) (8, 2320882) (9, 2374705)
        };
        
        \addplot+ [fill=BarBlue, color=BarBlue] coordinates {
        (0, 0) (1, 0) (2, 1333256) (3, 2503048) (4, 3230359) (5, 3827122) (6, 3280678) (7, 2481327) (8, 2180672) (9, 2251154)
        };
        
        \addplot+ [fill=BarPurple, color=BarPurple] coordinates {
          (0, 0) (1, 0) (2, 0) (3, 1884665) (4, 2911031) (5, 3398257) (6, 3621024) (7, 2989333) (8, 2423047) (9, 2050902)
        };
        
        \addplot+ [fill=BarYellow, color=BarYellow] coordinates {
          (0, 0) (1, 0) (2, 0) (3, 0) (4, 2410589) (5, 2832350) (6, 3143166) (7, 3185080) (8, 2769294) (9, 2174400)
        };
        
        \legend{$S=1$, $S=2$, $S=3$, $S=4$, $S=5$}
    
    \end{axis}
    \end{tikzpicture}
    \caption{Average quantities between 28--6--2021 and \\2--7--2021.}
\end{subfigure}
  \caption{Average outstanding queue size at distance $\delta=1,\dots,10$ ticks from the opposite best quote, shown separately for each spread size $S$.}
  \label{fig:average_q}
\end{figure}

\subsection{Estimating the Probability of a Mid-Price Change}\label{sec: results_midprice}

Since the analysis is symmetric for upward and downward mid-price movements, we focus without loss of generality on the probability of a mid-price increase. Model-implied probabilities are computed via the Laplace-transform representation in Proposition~\ref{proppricemove}, combined with numerical inversion using the COS method described in Section~\ref{sec: cosmethod}. 

To benchmark the model, we compute the empirical conditional probability of an upward mid-price move from the observed frequencies. The evaluation is performed on a five-day sample from 29--6--2021 to 5--7--2021. 

Specifically, let $t$ denote an event time and let $t_M$ be the first subsequent time at which the mid-price changes. The empirical probability of an increase, conditional on the state $(q_A,q_B,S)$, is defined as
\begin{equation}\label{emp_pricemove}
    P_{\mathrm{inc}}^{S}(q_A,q_B)
    =
    \frac{
    \#\big\{p_M(t_M)>p_M(t),\; Q_A(t)=q_A,\; Q_B(t)=q_B,\; S(t)=S\big\}
    }{
    \#\big\{p_M(t_M)\neq p_M(t),\; Q_A(t)=q_A,\; Q_B(t)=q_B,\; S(t)=S\big\}
    },
\end{equation}
where $\#$ denotes the number of occurrences in the data, $Q_A(t)$ and $Q_B(t)$ are the best-ask and best-bid queue sizes, and $S(t)$ is the spread.

\paragraph{Calibration windows.}
We consider two calibration schemes for the intensity parameters:
\begin{enumerate}
    \item \emph{Same weekday calibration:} parameters are estimated using the previous four occurrences of the same weekday (e.g., four previous Mondays to predict the next Monday).
    \item \emph{Rolling window calibration:} parameters are estimated using the preceding five trading days (Monday--Friday) to predict the following Monday.
\end{enumerate}
For each scheme, we compute model probabilities both with spread-independent and spread-dependent intensities.

\paragraph{Error metric.}
To assess predictive accuracy, we report the mean absolute percentage error (MAPE),
\begin{equation}
    \mathrm{MAPE}
    =
    \frac{1}{n}\sum_{k=1}^n 
    \left|\frac{P_k-\hat{P}_k}{P_k}\right|,
\end{equation}
where $P_k$ is the empirical probability, $\hat{P}_k$ the corresponding model estimate, and $n$ the total number of data points.

{\captionsetup{font=small}
\begin{table}[H]
    \centering
        \begin{adjustbox}{max width=\textwidth}
            \begin{tabular}{c r r r r}\toprule
            & \multicolumn{4}{c}{MAPE}\\ \cmidrule{2-5}
             & \multicolumn{2}{c}{Rolling window (preceding 5 days)}& \multicolumn{2}{c}{Same weekday (previous 4 instances)} \\ \cmidrule{2-5}
             $S$ & Spread independent rates & Spread dependent rates & Spread independent rates & Spread dependent rates \\ \cmidrule{1-5}
            1 & 11.3\% & \textcolor{black}{\textbf{10.4\%}} & 11.4\% & 10.6\% \\
            2 & 12.0\% & 11.4\% & 13.0\% & \textcolor{black}{\textbf{10.8}}\% \\
            3 & \textcolor{black}{\textbf{12.7\%}} & 13.3\% & 14.7\% & 14.6\% \\
            4 & 6.4\% & \textcolor{black}{\textbf{4.7\%}} & 6.6\% & 5.4\% \\
            5 & 10.8\% & 8.8\% & 10.8\% & \textcolor{black}{\textbf{8.1\%}} \\ \midrule
            Average & 10.6\% & \textcolor{black}{\textbf{9.7\%}} & 11.3\% & 9.9\% \\ \bottomrule
            \end{tabular}
        \end{adjustbox}
         \caption{MAPE between empirical and model-implied conditional probabilities of an upward mid-price move.}
         \label{tab: MAEMSE_pm}  
\end{table}}

Table~\ref{tab: MAEMSE_pm} shows that incorporating spread dependence generally improves predictive accuracy. Although no calibration scheme dominates for all spread sizes, the rolling-window calibration with spread-dependent intensities yields the lowest average MAPE, and is therefore used in the remainder of the numerical experiments.

Table~\ref{tab: prob_pricemove_week_before} shows empirical frequencies and model-implied probabilities on 5--7--2021 under the selected calibration. To ensure statistical reliability, we discard state configurations $(q_A,q_B,S)$ observed fewer than 100 times. Overall, the model reproduces the empirical monotonicity: the probability of an upward move decreases with $q_A$ and increases with $q_B$. The resulting MAPE values indicate that the proposed framework captures short-term mid-price dynamics reasonably well.

{\captionsetup{font=small}
\begin{table}[!htbp]
    \centering
        \begin{adjustbox}{max width=\textwidth}
            \begin{tabular}{c c r r r r r |r r r r r}\toprule
             & &\multicolumn{5}{c}{Empirical Probability}& \multicolumn{5}{c}{Model-Implied Probability}  \\ \cmidrule{3-12}
              & &\multicolumn{5}{c}{$q_A$}& \multicolumn{5}{c}{$q_A$}  \\ \cmidrule{3-12}
             & $q_B$ & 1 & 2 & 3 & 4 & 5 & 1 & 2 & 3 & 4 & 5 \\ \midrule
             & 1 & 50.3\% & 33.0\% & 22.9\% & 27.1\% & 22.4\% & 50.0\% & 34.7\% & 27.1\% & 22.6\% & 19.6\% \\
             & 2 &70.5\% & 56.6\% & - & - & - & 65.3\% & 50.0\% & 41.1\% & 35.3\% & 31.1\% \\
             $S=1$ & 3 & 78.7\% & - & - & - & - & 72.9\% & 58.9\% & 50.0\% & 43.8\% & 39.2\%\\
             & 4 & 78.2\% & - & - & - & - & 77.4\% & 64.7\% & 56.2\% & 50.0\% & 45.3\%\\
             & 5 & 81.4\% & - & - & - & - & 80.5\% & 68.9\% & 60.8\% & 54.8\% & 50.0\%\\ \midrule
             & 1 & 49.6\% & 38.5\% & 32.2\% & 25.2\% & 22.0\% & 50.0\% & 36.8\% & 30.9\% & 27.6\% & 25.6\%\\
             & 2 & 58.1\% & 48.5\% & 52.3\% & 45.5\% & 27.5\% & 63.3\% & 50.0\% & 43.2\% & 39.1\% & 36.4\%\\
             $S=2$ & 3 & 70.1\% & 49.9\% & - & - & - & 69.2\% & 56.8\% & 50.0\% & 45.7\% & 42.7\%\\
             & 4 & 75.1\% & 43.9\% & - & - & - & 72.4\% & 60.9\% & 54.4\% & 50.0\% & 47.0\%\\
             & 5 & 81.6\% & - & - & - & - & 74.5\% & 63.7\% & 57.3\% & 53.1\% & 50.0\%\\ \midrule
             & 1 & 49.9\% & 42.8\% & 39.7\% & 36.8\% & 31.8\% & 50.0\% & 38.9\% & 34.4\% & 32.2\% & 30.8\%\\
             & 2 & 54.7\% & 48.1\% & 52.0\% & 47.5\% & 52.6\% & 61.1\% & 50.0\% & 45.0\% & 42.2\% & 40.5\%\\
             $S=3$ & 3 & 56.3\% & 46.0\% & 45.3\% & 49.9\% & 72.2\% & 65.6\% & 55.0\% & 50.0\% & 47.12\% & 45.3\%\\
             & 4 & 62.5\% & 49.4\% & 59.7\% & 47.0\% & - & 67.8\% & 57.8\% & 52.9\% & 50.0\% & 48.1\%\\
             & 5 & 65.7\% & 51.9\% & - & - & - & 69.2\% & 59.5\% & 54.7\% & 51.9\% & 50.0\%\\ \midrule
             & 1 & 51.7\% & 44.0\% & 41.5\% & 38.6\% & 35.3\% & 50.0\% & 41.5\% & 38.8\% & 37.6\% & 37.0\%\\
             & 2 & 56.5\% & 47.0\% & 45.1\% & 43.5\% & 44.1\% & 58.5\% & 50.0\% & 47.0\% & 45.6\% & 44.9\%\\
             $S=4$ & 3 & 58.3\% & 52.8\% & 50.5\% & 52.8\% & 40.1\% & 61.2\% & 53.0\% & 50.0\% & 48.6\% & 47.8\%\\
             & 4 & 61.5\% & 54.8\% & 55.6\% & 51.7\% & 48.3\% & 62.4\% & 54.4\% & 51.4\% & 50.0\% & 49.2\%\\
             & 5 & 67.7\% & 58.7\% & 60.2\% & 54.0\% & 44.1\% & 63.0\% & 55.1\% & 52.2\% & 50.8\% & 50.0\%\\ \midrule
              & 1 & 39.1\% & 43.3\% & 49.3\% & 42.3\% & 51.0\% & 50.0\% & 44.4\% & 43.1\% & 42.7\% & 42.6\% \\
              & 2 & 56.8\% & 45.5\% & 66.8\% & 52.1\% & 40.0\% & 55.6\% & 50.0\% & 48.6\% & 48.2\% & 48.0\% \\
              $S=5 $& 3 & 63.2\% & 48.1\% & 53.9\% & 59.7\% & 53.7\% & 56.9\% & 51.4\% & 50.0\% & 49.5\% & 49.4\% \\
              & 4 & 65.0\% & 54.1\% & 39.4\% & 53.6\% & 50.7\% & 57.3\% & 51.8\% & 50.5\% & 50.0\% & 49.8\% \\
              & 5 & 65.7\% & 64.2\% & 42.5\% & 53.3\% & 45.2\% & 57.4\% & 52.0\% & 50.6\% & 50.2\% & 50.0\%\\
             \bottomrule
            \end{tabular}
        \end{adjustbox}
         \caption{Empirical probabilities and model-implied probabilities of an upward mid-price move on 5--7--2021, conditional on spread size $S$ and initial best-ask/best-bid queue sizes $(q_A,q_B)$.}
     \label{tab: prob_pricemove_week_before}  
\end{table}}

\subsection{Estimate the Fill Probability at the Best Quotes}\label{sec: fill_pro_result}

We next estimate the conditional probability that a limit order submitted at the best quote is executed before the mid-price moves. 
Model-implied fill probabilities are computed from Proposition~\ref{prop_fillprob}. 

Since the theoretical quantity is conditional on the event that the order is never cancelled, we construct an empirical proxy by focusing on orders that either (i) are filled before any mid-price change, or (ii) remain active until the mid-price moves and are cancelled only afterwards. 
We estimate
\begin{equation}\label{emp_fillprob}
    P_{\text{fill}}^S
    =
    \frac{
    \#\{\text{Fill},\; p_M(t_F)=p_M(t_0)\}
    }{
    \#\{\text{Fill},\; p_M(t_F)=p_M(t_0)\}
    +
    \#\{\text{Cancel},\; p_M(t_C)\neq p_M(t_0)\}
    },
\end{equation}
where $\#$ denotes the number of events, $t_0$ is the submission time, and $t_F$ and $t_C$ denote the execution and cancellation times, respectively. 
Here, $p_M(t)$ denotes the mid-price at time $t$. 
For readability, we omit the conditioning on $(Q_A(t_0),Q_B(t_0),S(t_0))=(q_A,q_B,S)$.

Model-implied probabilities are obtained by numerically inverting the Laplace transforms in Proposition~\ref{prop_fillprob} using the COS method. 
Table~\ref{tab: prob_fill_week_before} shows empirical and model-implied fill probabilities on 5--7--2021 for spread sizes $S\in\{1,\dots,5\}$. 
As before, we only report states observed at least $100$ times.

Overall, the model reproduces the main monotonicity patterns in the data. 
Performance is strongest for small queue sizes, while fill probabilities are slightly overestimated when the best-bid queue is large. 
This is unsurprising, since for $q_B>1$ the empirical execution probability is typically below $0.5\%$, making accurate estimation challenging. 
Nevertheless, the results indicate that the proposed framework provides a tractable and empirically reasonable approximation of short-term execution probabilities at the best quotes.

\begin{table}[!htbp]
    \centering
        \begin{adjustbox}{max width=\textwidth}
            \begin{tabular}{c c r r r r r |r r r r r}\toprule
             & &\multicolumn{5}{c}{Empirical Probability}& \multicolumn{5}{c}{Model-Implied Probability}  \\ \cmidrule{3-12}
              & &\multicolumn{5}{c}{$q_A$}& \multicolumn{5}{c}{$q_A$}  \\ \cmidrule{3-12}
             & $q_B$ & 1 & 2 & 3 & 4 & 5 & 1 & 2 & 3 & 4 & 5 \\ \midrule
             & 1 & 2.0\% & 5.3\% & 7.1\% & 13.4\% & - & 3.0\% & 3.9\% & 4.6\% & 5.1\% & 5.5\%  \\
             & 2 & 0.6\% & 0.0\% & -      & -      & - & 1.9\% & 2.7\% & 3.2\% & 3.7\% & 4.0\% \\
             $S=1$ & 3 & 0.3\% & -      & -      & -      & - & 1.5\% & 2.2\% & 2.6\% & 3.0\% & 3.3\%\\
             & 4 & 0.0\% & - & - & - & - & 1.2\% & 1.8\% & 2.2\% & 2.6\% & 2.9\%\\
             & 5 & 0.0\% & - & - & - & - & 0.9\% & 1.5\% & 1.9\% & 2.2\% & 2.5\%\\ \midrule
             & 1 & 1.3\% & 2.5\% & 4.1\% & 5.9\% & 5.9\% & 1.5\% & 1.8\% & 2.0\% & 2.1\% & 2.2\%\\
             & 2 & 0.3\% & 0.1\% & 0.0\% & -      & - & 1.0\% & 1.2\% & 1.3\% & 1.4\% & 1.5\%\\
             $S=2$ & 3 & 0.3\% & 1.6\% & -      & -      & - & 0.7\% & 0.9\% & 1.0\% & 1.1\% & 1.2\%\\
             & 4 & 0.2\% & -      & -      & -      & - & 0.6\% & 0.8\% & 0.9\% & 1.0\% & 1.0\%\\
             & 5 & 0.0\% & 0.0\% & 0.0\% & - & - & 0.5\% & 0.7\% & 0.8\% & 0.8\% & 0.9\%\\ \midrule
             & 1 & 0.5\% & 0.9\% & 1.4\% & 1.1\% & 0.8\% & 1.2\% & 1.3\% & 1.4\% & 1.4\% & 1.4\%\\
             & 2 & 0.1\% & 0.0\% & 0.3\% & 0.0\% & 0.0\% & 0.8\% & 0.9\% & 1.0\% & 1.0\% & 1.0\%\\
             $S=3$ & 3 & 0.1\% & 0.1\% & 0.0\% & 0.0\% & - & 0.7\% & 0.7\% & 0.8\% & 0.8\% & 0.8\%\\
             & 4 & 0.1\% & 0.0\% & 0.0\% & -  & - & 0.5\% & 0.5\% & 0.5\% & 0.6\% & 0.7\%\\
             & 5 & 0.1\% & 0.0\% & -      & -  & - & 0.5\% & 0.5\% & 0.6\% & 0.6\% & 0.6\%\\ \midrule
             & 1 & 1.5\% & 0.5\% & 0.3\% & 0.5\% & 0.3\% & 1.1\% & 1.2\% & 1.2\% & 1.2\% & 1.2\%\\
             & 2 & 0.0\% & 0.0\% & 0.2\% & 0.1\% & 0.0\% & 0.8\% & 0.8\% & 0.8\% & 0.8\% & 0.8\%\\
             $S=4$ & 3 &0.0\% & 0.1\% & 0.0\% & 0.0\% & 0.0\% & 0.6\% & 0.7\% & 0.7\% & 0.7\% & 0.7\%\\
             & 4 & 0.1\% & 0.0\% & 0.0\% & 0.0\% & 0.0\% & 0.5\% & 0.6\% & 0.6\% & 0.6\% & 0.6\%\\
             & 5 & 0.0\% & 0.0\% & 0.0\% & 0.0\% & - & 0.5\% & 0.5\% & 0.5\% & 0.5\% & 0.5\%\\\midrule
             & 1 & 0.0\% & - & - & - & - & 1.1\% & 1.1\% & 1.1\% & 1.1\% & 1.1\%\\
             & 2 & 0.0\% & - & - & - & - & 0.6\% & 0.7\% & 0.7\% & 0.7\% & 0.7\%\\
             $S=5$ & 3 & - & - & - & - & - & 0.5\% & 0.5\% & 0.5\% & 0.7\% & 0.5\%\\
             & 4 & - & - & - & - & - & 0.3\% & 0.3\% & 0.4\% & 0.4\% & 0.4\%\\
             & 5 & - & - & - & - & - & 0.3\% & 0.3\% & 0.3\% & 0.3\% & 0.3\%\\\bottomrule
            \end{tabular}
        \end{adjustbox}
        \caption{Empirical and model-implied fill probabilities of a best-bid order on 5--7--2021, conditional on spread size $S$ and initial best-ask/best-bid queue sizes $(q_A,q_B)$.}

     \label{tab: prob_fill_week_before}    
\end{table}

\subsection{Estimate the Fill Probability at a Price Deeper than the Best Quotes}\label{prop_level2_result}

In this section, we estimate the fill probability of a bid order posted one tick below the best bid, i.e., at $p_{B-}=p_B-1$. 
Recall that $W_{B-}(t)$ denotes the number of remaining orders at level $p_{B-}$ at time $t$ originating from the initial queue $Q_{B-}(0)$.

As observed in Section~\ref{sec: fill_pro_result}, empirical fill probabilities become negligible when $W_{B-}>2$. 
We therefore restrict attention to the cases $m\in\{1,2\}$ and assume that for $m>2$,
\begin{equation}
     \mathbb{P}\!\left[\epsilon_{B-}<\tau^B \,\middle|\, W_{B-}(\tau_B^\text{quote})=m,\,
     Q_{A}(\tau_B^\text{quote})=n\right]=0.  
\end{equation}

To estimate the probability in \eqref{prob_level2}, we specialize to $(i,j)=(B,A)$, which yields
\begin{equation}\label{prob_level2_numerical}
\resizebox{1.0\hsize}{!}{$
    \begin{aligned}
        &\mathbb{P}[\tau_B^\text{quote}<\tau_B^\text{other}]\cdot\\
        &\left(\sum_{m=1}^{q_0^{B-}}\sum_{n=1}^{N_A}\bigg(\mathbb{P}[\epsilon_{B-}<\tau^B \mid W_{B-}(\tau_B^\text{quote})=m,Q_{A}(\tau_B^\text{quote})=n]\cdot\mathbb{P}[W_{B-}(\tau_B^\text{quote})=m]\cdot\mathbb{P}[Q_{A}(\tau_B^\text{quote})=n]\bigg)\right).
    \end{aligned}
$}    
\end{equation}

We focus on the cases where the spread at submission equals one or two ticks, since Model~III achieved its best performance for these spread sizes in Section~\ref{sec: fill_pro_result}. 

Table~\ref{tab: quantity_ask_pricemove} shows the empirical distribution of the best-ask queue size after a downward move of the best bid. 
For both $S=1$ and $S=2$, the probability mass is concentrated on $q_A\in\{1,2\}$. 
We therefore truncate the distribution by setting $\mathbb{P}[Q_A(\tau_B^\text{quote})=n]=0$ for $n>2$ and take $N_A=2$. 
Following Cont and De Larrard (2013), we use the empirical frequencies in Table~\ref{tab: quantity_ask_pricemove} as estimates of $\mathbb{P}[Q_A(\tau_B^\text{quote})=n]$ for $n=1,2$. 
Under this truncation, we set $q_0^{B-}=N_A=2$ in \eqref{prob_level2_numerical}.

\begin{table}[h]
    \centering
        \begin{adjustbox}{max width=\textwidth}
            \begin{tabular}{c r r r r r r}\toprule
            &\multicolumn{6}{c}{$q_A$}\\ \cmidrule{2-7}
              $S$ & 1 & 2 & 3 & 4 & 5 & $>5$ \\ \midrule
              1 & 76.2\% & 22.0\% & 1.6\% & 0.0\% & 0.0\% & 0.2\% \\
              2 & 82.2\% & 15.2\% & 1.4\% & 0.0\% & 0.0\% & 0.6\% \\\bottomrule
            \end{tabular}
        \end{adjustbox}
        \caption{Empirical distribution of the best-ask queue size $q_A$ after a downward move of the best bid, conditional on spread size $S$, using data from 29--6--2021 to 5--7--2021.}   
        \label{tab: quantity_ask_pricemove}
 \end{table}

We next consider the case where $S=1$ at submission, so that $S(\tau_B^\text{quote})=S+1$ after the bid moves down. 
Model-implied probabilities are computed using order book data from 21--6--2021 to 2--7--2021. 
Table~\ref{tab: prob_level2_result} shows the corresponding empirical and model-implied fill probabilities. 
Even with two weeks of data, some queue-size configurations occur fewer than 100 times and are therefore excluded from the empirical estimates.

Overall, the model captures the qualitative dependence of fill probabilities at the second best quote on prevailing queue sizes, with a mild tendency to overestimate execution. 
This is not unexpected, since the relevant probabilities are very small and the computation may accumulate estimation errors from both best-quote fill probabilities and mid-price move probabilities. 
Nevertheless, the results demonstrate that the proposed framework remains tractable beyond the best quotes and provides model-implied fill probabilities that are broadly consistent with empirical patterns observed in the FX spot market.

\begin{table}[!htbp]
         \centering
        \begin{adjustbox}{max width=\textwidth}
            \begin{tabular}{c c r r r r r |r r r r r}\toprule
             & &\multicolumn{5}{c}{Empirical Probability}& \multicolumn{5}{c}{Model-Implied Probability}  \\ \cmidrule{3-12}
              & &\multicolumn{5}{c}{$q_A$}& \multicolumn{5}{c}{$q_A$}  \\ \cmidrule{3-12}
             & $q_B$ & 1 & 2 & 3 & 4 & 5 & 1 & 2 & 3 & 4 & 5 \\ \midrule
             \multirow{4}{*}{$q_{B-}=1$}& 1 & 0.19\% & 0.27\% & 0.68\% & 2.16\% & - & 0.75\% & 0.97\% & 1.09\% & 1.16\% & 1.21\%  \\
             & 2 & 0.23\% & 0.25\% & 0.39\% & 0.67\% & - & 0.53\% & 0.75\% & 0.88\% & 0.97\% & 1.03\% \\
              & 3 & 0.11\% & 0.69\% & -     & -     & - & 0.42\% & 0.62\% & 0.75\% & 0.84\% & 0.91\%\\ 
             & 4 & 0.12\% & -     & -     & -     & - & 0.35\% & 0.54\% & 0.66\% & 0.75\% & 0.82\%  \\ \midrule
             \multirow{4}{*}{$q_{B-}=2$} & 1 & 0.06\% & 0.44\% & -     & -     & - & 0.63\% & 0.81\% & 0.91\% & 0.96\% & 1.00\% \\
             & 2 & 0.12\% & 0.37\% & 0.00\% & -     & - & 0.47\% & 0.67\% & 0.78\% & 0.86\% & 0.91\% \\
             & 3 & 0.12\% & 0.00\% & -     & -     & - & 0.38\% & 0.57\% & 0.69\% & 0.77\% & 0.83\% \\
             & 4 & 0.00\% & -     & -     & -     & - & 0.32\% & 0.50\% & 0.61\% & 0.70\% & 0.76\% \\\midrule
             \multirow{4}{*}{$q_{B-}=3$} & 1 & 0.28\% & 1.12\% & -     & -     & - & 0.42\% & 0.54\% & 0.60\% & 0.64\% & 0.67\% \\
             & 2 & 0.09\% & 0.22\% & -     & -     & - & 0.38\% & 0.53\% & 0.62\% & 0.68\% & 0.72\% \\
             & 3 & 0.04\% & -     & -     & -     & - & 0.33\% & 0.49\% & 0.59\% & 0.66\% & 0.71\% \\
             & 4 & 0.00\% & -     & -     & -     & - & 0.29\% & 0.44\% & 0.55\% & 0.62\% & 0.68\% \\\midrule
             \multirow{4}{*}{$q_{B-}=4$} & 1 & 0.06\% & 0.71\% & -     & -     & - & 0.33\% & 0.42\% & 0.47\% & 0.50\% & 0.53\% \\
             & 2 & 0.08\% & 0.62\% & -     & -     & - & 0.32\% & 0.45\% & 0.52\% & 0.58\% & 0.61\% \\
             & 3 & 0.00\% & -     & -     & -     & - & 0.29\% & 0.43\% & 0.52\% & 0.58\% & 0.63\% \\
             & 4 & 0.03\% & -     & -     & -     & - & 0.26\% & 0.40\% & 0.49\% & 0.56\% & 0.61\% \\\midrule
             \multirow{4}{*}{$q_{B-}=5$} & 1 & 0.07\% & 0.00\% & -     & -     & - & 0.27\% & 0.35\% & 0.39\% & 0.42\% & 0.43\% \\
             & 2 & 0.08\% & 0.89\% & -     & -     & - & 0.28\% & 0.39\% & 0.46\% & 0.50\% & 0.53\%  \\
             & 3 & 0.00\% & -     & -     & -     & - & 0.26\% & 0.38\% & 0.46\% & 0.52\% & 0.56\% \\
             & 4 & 0.03\% & -     & -     & -     & - & 0.24\% & 0.36\% & 0.45\% & 0.51\% & 0.56\% \\\bottomrule
            \end{tabular}
        \end{adjustbox}
        \caption{Empirical and model-implied fill probabilities for orders posted at $p_B-1$ with $S=1$, using data from 21--6--2021 to 2--7--2021.}

     \label{tab: prob_level2_result}
\end{table}

\section{Conclusion}\label{conclusion}

This paper develops a tractable framework for computing fill probabilities of limit orders posted at different depths in a limit order book. We model the order book as a collection of interacting queueing systems with state-dependent order arrivals and cancellations, allowing the intensities to depend on both queue sizes and additional stylized market factors. Within this general setting, we derive semi-analytical expressions for key execution-related probabilities, including mid-price move probabilities and fill probabilities at the best quotes and one level deeper. 

We also show that several existing models arise as special cases of our framework, and we illustrate the practical performance of the proposed methodology using FX spot market data. The numerical results indicate that the derived formulas remain computationally tractable and produce fill probability estimates that align well with empirical patterns. Further refinements of the stochastic intensity specifications may improve quantitative accuracy for specific markets, but such extensions are left for future work.
\\

\textbf{Acknowledgment}: The authors thank MN for providing foreign exchange spot market limit order book data used in the empirical experiments.

\appendix

\section{Numerical Methods for Inverting a Laplace Transform}\label{numericalappendix}

\subsection{Euler Method}\label{appendix:euler_method}

We briefly summarize the Euler method of Abate and Whitt~(1995) for numerically inverting Laplace transforms. 
This method is based on a Fourier-series representation of the Bromwich integral and applies Euler summation to accelerate convergence.

Let $\hat{f}(s)$ denote the Laplace transform of a function $f(t)$. The Bromwich inversion formula gives
\begin{equation}
    f(t)
    =
    \frac{1}{2\pi i}
    \int_{\gamma-i\infty}^{\gamma+i\infty}
    e^{st}\hat{f}(s)\,\mathrm{d}s,
\end{equation}
where $\gamma>0$ is chosen so that the contour lies to the right of all singularities of $\hat{f}$.
Following Abate and Whitt~(1995), this integral can be rewritten as
\begin{equation}
    f(t)
    =
    \frac{2e^{\gamma t}}{\pi}
    \int_0^\infty
    \Re\!\left(\hat{f}(\gamma+iu)\right)\cos(ut)\,\mathrm{d}u,
\end{equation}
where $\Re(z)$ denotes the real part of $z\in\mathbb{C}$.

Approximating the integral using the trapezoidal rule with step size $h>0$ yields
\begin{equation}
    f(t)
    \approx
    f_h(t)
    :=
    \frac{he^{\gamma t}}{\pi}
    \Re\!\left(\hat{f}(\gamma)\right)
    +
    \frac{2he^{\gamma t}}{\pi}
    \sum_{k=1}^{\infty}
    \Re\!\left(\hat{f}(\gamma+ikh)\right)\cos(kht).
\end{equation}
A convenient choice is $h=\frac{\pi}{2t}$ and $\gamma=\frac{A}{2t}$, which gives the classical Euler inversion formula
\begin{equation}\label{EulerAbate}
    f_h(t)
    =
    \frac{e^{A/2}}{2t}
    \Re\!\left\{
    \hat{f}\!\left(\frac{A}{2t}\right)
    \right\}
    +
    \frac{e^{A/2}}{t}
    \sum_{k=1}^{\infty}
    (-1)^k
    \Re\!\left\{
    \hat{f}\!\left(\frac{A+2k\pi i}{2t}\right)
    \right\}.
\end{equation}

If $|f(t)|\leq 1$ for all $t$ (as is the case for cumulative distribution functions), the discretization error satisfies
\begin{equation}
    \bigl|f(t)-f_h(t)\bigr|
    \leq
    \frac{e^{-A}}{1-e^{-A}}.
\end{equation}
For small $e^{-A}$, this bound is approximately $e^{-A}$. Hence, to achieve a target accuracy of at most $10^{-\gamma}$, one may choose $A=\gamma\log(10)$. A commonly used value is $\gamma=8$, leading to $A\approx 18.4$.

To compute~\eqref{EulerAbate} efficiently, Abate and Whitt~(1995) recommend Euler summation, which forms a weighted average of the last $m$ partial sums using binomial weights. Define the $n$-th partial sum by
\begin{equation}
    s_n(t)
    :=
    \frac{e^{A/2}}{2t}
    \Re\!\left\{
    \hat{f}\!\left(\frac{A}{2t}\right)
    \right\}
    +
    \frac{e^{A/2}}{t}
    \sum_{k=1}^{n}
    (-1)^k
    \Re\!\left\{
    \hat{f}\!\left(\frac{A+2k\pi i}{2t}\right)
    \right\}.
\end{equation}
Then the Euler-accelerated approximation is given by
\begin{equation}
    f(t)
    \approx
    E(m,n,t)
    :=
    \sum_{k=0}^{m}
    \binom{m}{k}2^{-m}
    s_{n+k}(t).
\end{equation}

In practice, Abate and Whitt~(1995) suggest choosing $m=11$ and $n=15$, and increasing $n$ if higher accuracy is required. We refer the reader to the original paper for further details and implementation considerations.

\subsection{COS Method}\label{sec: cosmethod}

In addition to the Euler method, Laplace transforms can be inverted numerically using Fourier-based techniques.
In particular, by evaluating the Laplace transform along the imaginary axis, one obtains the characteristic function.
Indeed, setting $s=i\omega$ yields
\begin{equation}
    \hat{f}(i\omega)
    =
    \int_{0}^{\infty} e^{-i\omega t} f(t)\,\mathrm{d}t,
\end{equation}
which corresponds to the Fourier transform of $f$.
This observation motivates the COS method of Fang and Oosterlee~(2009), which approximates the target function via a truncated Fourier--cosine expansion.
The COS method typically achieves exponential convergence for smooth densities and has linear computational complexity in the number of expansion terms.

\medskip

\noindent
\textbf{Fourier--cosine expansion.}
Let $f(t)$ be supported on a finite interval $[a,b]\subset\mathbb{R}$. Its Fourier--cosine expansion is given by
\begin{equation}\label{sumcos}
    f(t)
    =
    \sideset{}{'}\sum_{k=0}^{\infty}
    \bar{A}_k
    \cos\!\left(k\pi\frac{t-a}{b-a}\right),
\end{equation}
where the coefficients satisfy
\begin{equation}\label{Ak}
    \bar{A}_k
    =
    \frac{2}{b-a}
    \int_{a}^{b}
    f(t)
    \cos\!\left(k\pi\frac{t-a}{b-a}\right)\mathrm{d}t,
\end{equation}
and the notation $\sum'$ indicates that the $k=0$ term is weighted by $\frac{1}{2}$.

\medskip

\noindent
\textbf{Truncation of the Laplace integral.}
Since the Bromwich inversion integral requires the Fourier transform of $f$, it is natural to approximate the infinite-domain integral by truncating the support.
Choosing $[a,b]$ such that the tail mass outside the interval is negligible, define the truncated transform
\begin{equation}\label{FT}
    \hat{f}^{*}(s)
    :=
    \int_{a}^{b} e^{-st}f(t)\,\mathrm{d}t,
\end{equation}
so that $\hat{f}^{*}(s)\approx \hat{f}(s)$.

\medskip

\noindent
\textbf{Coefficient approximation using the Laplace transform.}
Substituting $\omega=-\frac{k\pi}{b-a}$ and multiplying by $e^{-i\frac{ak\pi}{b-a}}$, we obtain
\begin{equation}
    \hat{f}^{*}\!\left(-i\frac{k\pi}{b-a}\right)
    \exp\!\left(-i\frac{ak\pi}{b-a}\right)
    =
    \int_{a}^{b}
    \exp\!\left(i k\pi\frac{t-a}{b-a}\right)
    f(t)\,\mathrm{d}t.
\end{equation}
Taking the real part and using $e^{iu}=\cos(u)+i\sin(u)$ gives
\begin{equation}\label{realpart}
    \Re\!\left\{
    \hat{f}^{*}\!\left(-i\frac{k\pi}{b-a}\right)
    \exp\!\left(-i\frac{ak\pi}{b-a}\right)
    \right\}
    =
    \int_{a}^{b}
    f(t)\cos\!\left(k\pi\frac{t-a}{b-a}\right)\mathrm{d}t.
\end{equation}
Comparing with~\eqref{Ak}, the coefficients can therefore be approximated by
\begin{equation}
    \bar{A}_k
    =
    \frac{2}{b-a}
    \Re\!\left\{
    \hat{f}^{*}\!\left(-i\frac{k\pi}{b-a}\right)
    \exp\!\left(-i\frac{ak\pi}{b-a}\right)
    \right\}
    \approx
    \bar{F}_k,
\end{equation}
where
\begin{equation}
    \bar{F}_k
    :=
    \frac{2}{b-a}
    \Re\!\left\{
    \hat{f}\!\left(-i\frac{k\pi}{b-a}\right)
    \exp\!\left(-i\frac{ak\pi}{b-a}\right)
    \right\}.
\end{equation}
Hence, truncating the cosine expansion at $N$ terms yields the COS approximation
\begin{equation}\label{COS_approx}
    f(t)
    \approx
    f_{N}(t)
    :=
    \sideset{}{'}\sum_{k=0}^{N-1}
    \bar{F}_k
    \cos\!\left(k\pi\frac{t-a}{b-a}\right).
\end{equation}

\medskip

\noindent
\textbf{Choice of truncation interval.}
A crucial step in the COS method is selecting the integration range $[a,b]$.
Fang and Oosterlee~(2009) propose choosing
\begin{equation}
    [a,b]
    =
    \left[
    c_1 - L\sqrt{c_2+\sqrt{c_4}},
    \;
    c_1 + L\sqrt{c_2+\sqrt{c_4}}
    \right],
\end{equation}
where $L\in[6,12]$ and $c_n$ denotes the $n$-th cumulant of the underlying distribution.
These cumulants can be obtained from the cumulant generating function
\begin{equation}
    C_X(t)
    :=
    \log\mathbb{E}[e^{tX}]
    =
    \log \hat{f}(-t),
\end{equation}
so that
\begin{equation}
    c_n
    =
    \left.
    \frac{\mathrm{d}^n}{\mathrm{d}t^n}C_X(t)
    \right|_{t=0},
    \qquad n\geq 1.
\end{equation}
In particular,
\begin{equation}
    c_1
    =
    \left.\frac{\mathrm{d}}{\mathrm{d}t}C_X(t)\right|_{t=0},
    \qquad
    c_2
    =
    \left.\frac{\mathrm{d}^2}{\mathrm{d}t^2}C_X(t)\right|_{t=0},
    \qquad
    c_4
    =
    \left.\frac{\mathrm{d}^4}{\mathrm{d}t^4}C_X(t)\right|_{t=0}.
\end{equation}

For further details and practical implementation aspects of the COS method, we refer to Fang and Oosterlee~(2009).

\section{Numerical Methods for Continued Fractions}\label{numerical2appendix}

In this appendix, we briefly describe a stable numerical procedure for evaluating continued fractions, which arise in the Laplace transform expressions used throughout the paper.

Consider a continued fraction of the form
\begin{equation}\label{Eq: LentzCF}
    f
    =
    \frac{a_1}{b_1+}
    \frac{a_2}{b_2+}
    \frac{a_3}{b_3+}\cdots,
\end{equation}
where $\{a_k\}_{k\geq 1}$ and $\{b_k\}_{k\geq 1}$ are given sequences. 
The $k$-th approximant (or convergent) of $f$ is defined by truncating~\eqref{Eq: LentzCF} after $k$ terms:
\begin{equation}
    f^{(k)}
    =
    \frac{a_1}{b_1+}
    \frac{a_2}{b_2+}
    \cdots
    \frac{a_k}{b_k}
    =
    \frac{A_k}{B_k}.
\end{equation}
The numerator and denominator sequences $\{A_k\}$ and $\{B_k\}$ satisfy the same second-order recurrence relation
\begin{equation}\label{recurrence}
    \begin{cases}
        A_k = b_k A_{k-1} + a_k A_{k-2},\\[0.3em]
        B_k = b_k B_{k-1} + a_k B_{k-2},
    \end{cases}
\end{equation}
with initial conditions
\[
A_0=0,\qquad A_1=a_1,\qquad B_0=1,\qquad B_1=b_1.
\]

A straightforward approximation of $f$ is obtained by computing $f^{(k)}$ for a fixed truncation level $k$. 
However, as discussed by Crawford and Suchard~(2012), selecting $k$ in advance may lead to numerical instability or unnecessary computational cost. 
Instead, we adopt the modified Lentz method, originally proposed by Thompson and Barnett~(1986) and Press and Teukolsky~(1988), which provides a stable iterative scheme for evaluating~\eqref{Eq: LentzCF}.

\medskip

\noindent
\textbf{Modified Lentz method.}
Define the ratios
\begin{equation}
    C_k := \frac{A_k}{A_{k-1}},
    \qquad
    D_k := \frac{B_{k-1}}{B_k}.
\end{equation}
Using the recurrence relations~\eqref{recurrence}, these quantities satisfy
\begin{equation}
    \begin{cases}
        C_k = b_k + \dfrac{a_k}{C_{k-1}},\\[1em]
        D_k = \dfrac{1}{b_k + a_k D_{k-1}}.
    \end{cases}
\end{equation}
The approximants can then be updated recursively via
\begin{equation}
    f^{(k)} = f^{(k-1)}\,C_k D_k.
\end{equation}
The iteration is terminated once the relative update is sufficiently small, i.e.,
\begin{equation}
    \bigl|C_k D_k - 1\bigr| < \varepsilon,
\end{equation}
for a prescribed tolerance $\varepsilon>0$. Further discussion on convergence properties and numerical considerations can be found in Lorentzen and Waadeland~(2008).

\section{Proofs}\label{proofappendix}
\subsection{Lemma~\ref{LTminofexponential}}

\begin{proof}
Let $X\sim \mathrm{Exp}(\Lambda)$ with $\Lambda>0$, and assume that $X$ is independent of $Y$. 
For $t\geq 0$, we have
\begin{equation}
    \mathbb{P}[X\wedge Y>t]
    =
    \mathbb{P}[X>t]\mathbb{P}[Y>t]
    =
    e^{-\Lambda t}\bigl(1-F_Y(t)\bigr),
\end{equation}
where $F_Y$ denotes the cumulative distribution function of $Y$. Hence,
\begin{equation}
    F_{X\wedge Y}(t)
    =
    \mathbb{P}[X\wedge Y\leq t]
    =
    1-e^{-\Lambda t}\bigl(1-F_Y(t)\bigr),
    \qquad t\geq 0.
\end{equation}
Differentiating yields the density of $X\wedge Y$:
\begin{equation}
    f_{X\wedge Y}(t)
    =
    \frac{\mathrm{d}}{\mathrm{d}t}F_{X\wedge Y}(t)
    =
    e^{-\Lambda t}\Bigl(f_Y(t)+\Lambda\bigl(1-F_Y(t)\bigr)\Bigr),
    \qquad t\geq 0.
\end{equation}

Taking Laplace transforms, we obtain
\begin{equation}
\begin{aligned}
    \hat{f}_{X\wedge Y}(s)
    &=
    \int_{0}^{\infty} e^{-st} f_{X\wedge Y}(t)\,\mathrm{d}t\\
    &=
    \int_{0}^{\infty} e^{-(s+\Lambda)t}f_Y(t)\,\mathrm{d}t
    +
    \Lambda
    \int_{0}^{\infty} e^{-(s+\Lambda)t}\bigl(1-F_Y(t)\bigr)\,\mathrm{d}t\\
    &=
    \hat{f}_Y(s+\Lambda)
    +
    \Lambda
    \int_{0}^{\infty} e^{-(s+\Lambda)t}\bigl(1-F_Y(t)\bigr)\,\mathrm{d}t.
\end{aligned}
\end{equation}

To evaluate the remaining integral, we integrate by parts:
\begin{equation}
\begin{aligned}
    \int_{0}^{\infty} e^{-(s+\Lambda)t}\bigl(1-F_Y(t)\bigr)\,\mathrm{d}t
    &=
    \left[
    -\frac{1}{s+\Lambda}e^{-(s+\Lambda)t}\bigl(1-F_Y(t)\bigr)
    \right]_{0}^{\infty}
    -
    \frac{1}{s+\Lambda}
    \int_{0}^{\infty} e^{-(s+\Lambda)t} f_Y(t)\,\mathrm{d}t\\
    &=
    \frac{1}{s+\Lambda}
    -
    \frac{1}{s+\Lambda}\hat{f}_Y(s+\Lambda).
\end{aligned}
\end{equation}
Therefore,
\begin{equation}
\begin{aligned}
    \hat{f}_{X\wedge Y}(s)
    &=
    \hat{f}_Y(s+\Lambda)
    +
    \Lambda\left(
    \frac{1}{s+\Lambda}
    -
    \frac{1}{s+\Lambda}\hat{f}_Y(s+\Lambda)
    \right)\\
    &=
    \hat{f}_Y(s+\Lambda)
    +
    \frac{\Lambda}{s+\Lambda}
    \Bigl(1-\hat{f}_Y(s+\Lambda)\Bigr).
\end{aligned}
\end{equation}
\end{proof}

\bibliographystyle{plain} 
\bibliography{refs} 
\nocite{*}

\end{document}